\documentclass{svmult}

\usepackage{mathptmx}       
\usepackage{helvet}         
\usepackage{courier}        
\usepackage{type1cm}        
\usepackage{amsmath}        
\usepackage{makeidx}         
\usepackage{graphicx}        
\usepackage{multicol}        
\usepackage[bottom]{footmisc}

\usepackage{tikz}
\usetikzlibrary{decorations.pathmorphing}
\usetikzlibrary{backgrounds}

\makeindex             

\begin{document}

\title*{Heat transport in low dimensions: introduction and phenomenology}
\titlerunning{Heat transport in low dimensions}
\author{Stefano Lepri, Roberto Livi and Antonio Politi}
\institute{Stefano Lepri \at Consiglio Nazionale delle Ricerche, Istituto dei Sistemi Complessi, 
via Madonna del Piano 10, I-50019 Sesto Fiorentino, Italy, \email{stefano.lepri@isc.cnr.it}
\and Roberto Livi \at Dipartimento di Fisica e Astronomia and CSDC, Universit\`a di Firenze, 
via G. Sansone 1 I-50019, Sesto Fiorentino, Italy,   \email{roberto.livi@unifi.it} 
\and Antonio Politi \at Institute for Complex Systems and Mathematical Biology \& SUPA
University of Aberdeen, Aberdeen AB24 3UE, United Kingdom, \email{a.politi@abdn.ac.uk}}
%
%
\maketitle

\abstract*{In this chapter we introduce some of the basic models
and concepts that will be discussed throughout the volume. 
In particular we describe systems of nonlinear oscillators
arranged on low-dimensional lattices and summarize 
the phenomenology of their transport properties.}

\abstract{In this chapter we introduce some of the basic models
and concepts that will be discussed throughout the volume. 
In particular we describe systems of nonlinear oscillators
arranged on low-dimensional lattices and summarize
the phenomenology of their transport properties.}

\section{Introduction}

In this first chapter we review the main properties of low-dimensional lattices 
of coupled classical oscillators. We will describe how
reduced dimensionality and conservation laws conspire in giving rise to
unusual relaxation and transport properties. 
The aim is to provide both a general introduction to the general phenomenology
and to guide the reader in the volume reading (where appropriate we indeed point to 
the more detailed analyses developed in the subsequent chapters).

\begin{figure}[t]
\begin{center}
\includegraphics[width=6cm]{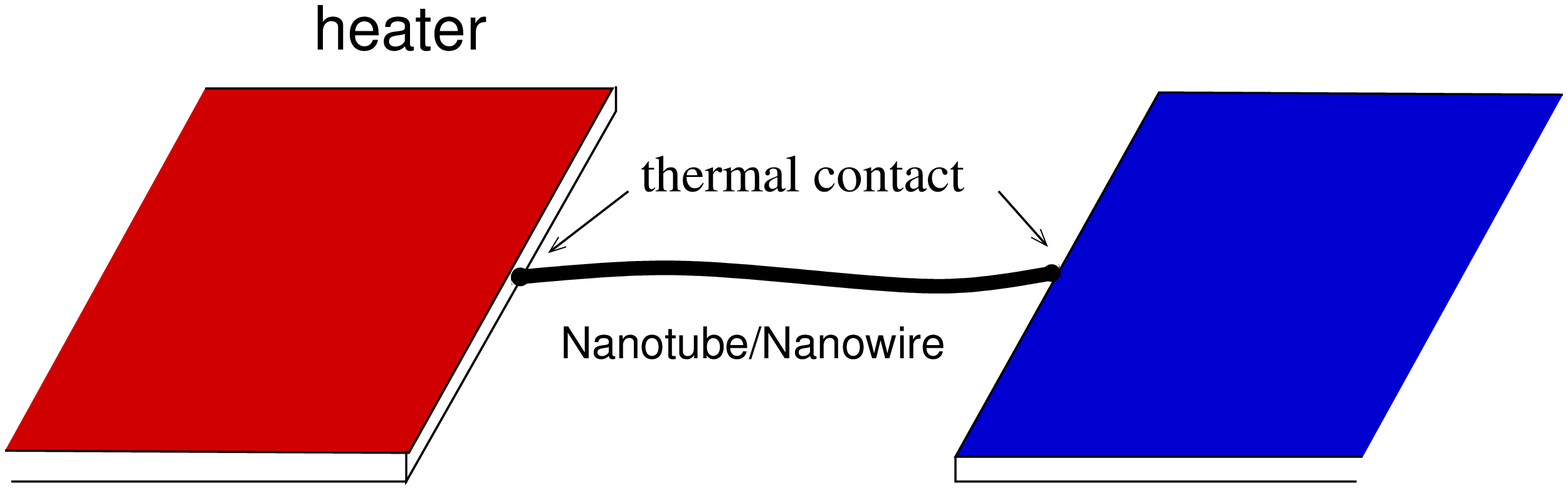}
\includegraphics[width=4cm]{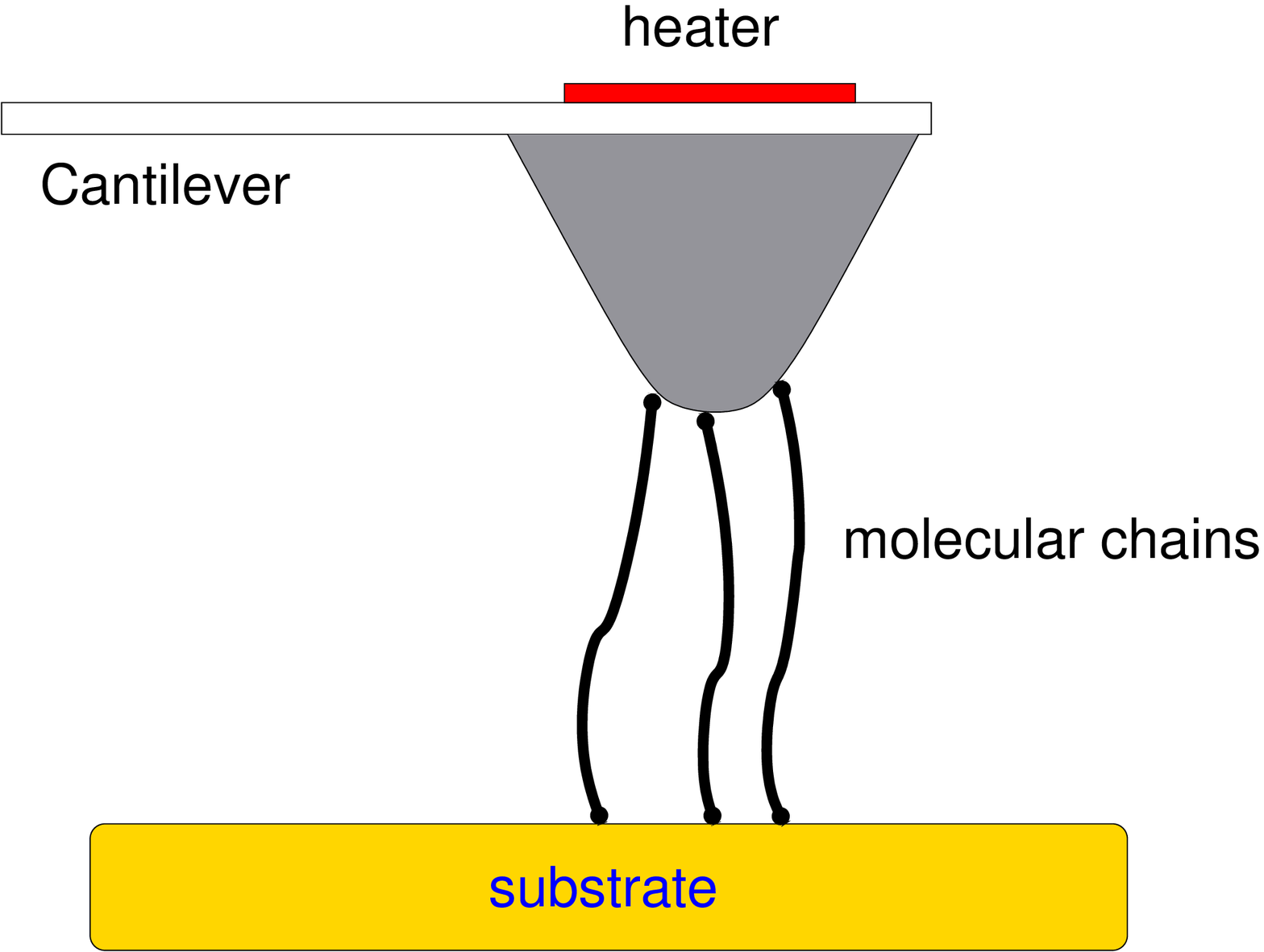}
\caption{Sketch of two experimental setups illustrating the physical setting.
Left: a nanotube or nanowire suspended between two contacts acting as heaters
and probes, see \cite{Chang08} and Chapter 8. Right: a scanning thermal microscopy setup whereby an assembly
of molecular chains with one end attached on a substrate is heated through a cantilever
tip \cite{Meier2014}.  }
\end{center}
\label{fig:setups}       
\end{figure}

For the sake of concreteness, one may think of quasi-1D objects, like 
long molecular chains or nanowires, suspended between two contacts which
play the role of thermal reservoirs.
Examples such experimental setups that will be repeatedly discussed throughout the volume
are schematically depicted in Fig.~\ref{fig:setups}.  

We start section~\ref{sec:models} by introducing the main models without technicalities and providing
the relevant definitions.
Section~\ref{sec:signatures} contains a summary of the different properties
that is worth testing to characterize heat transport in a physical system. The natural
starting point is the effective conductivity in finite systems, which diverges with the
system-size in the case of anomalous transport. The existence of long-time tails in the 
equilibrium correlation functions is another way of probing the system dynamics, together with the
diffusion of localized perturbations and the relaxation of spontaneous fluctuations.
Another, not much explored property, is the shape of the temperature profile that is
strictly nonlinear even in the limit of small temperature differences, when heat
transport is anomalous.

In section~\ref{sec:universality}, we present the overall scenario, making reference
to the universality classes unveiled by the various theoretical
approaches. More specifically, we emphasize the relationship with the evolution of rough
interfaces and thereby the Kardar-Parisi-Zhang equation. 
Coupled rotors represent an important subclass of 1D systems where heat conduction is normal in spite
of momentum conservation: their behaviour is reviewed in Sec.~\ref{sec:rotors}.

The expected scenario in two-dimensions (namely the logarithmic divergence of heat
conductivity) is discussed in Sec.~\ref{sec:2D}, while the peculiar behaviour of
integrable systems is briefly reviewed in Sec.~\ref{sec:integrable}.
In section~\ref{sec:coupled-transport}, we discuss the more general physical setup, where
another quantity is being transported besides energy. This is the problem of coupled transport,
where the interaction between the two processes may give rise to unexpected phenomena
even when the transport is altogether normal. In particular, we consider
a chain of coupled rotors in the presence of an additional torque, where the second quantity is
angular momentum and the discrete nonlinear Schr\"odinger equation, where the second quantity
is the norm (or mass). 
Finally, the still open problems are recalled in section~\ref{sec:conclusions}.

\section{Models}
\label{sec:models}

The simplest microscopic dynamical model for the characterization of heat conduction 
consists of a chain of $N$ classical point--like particles with mass $m$ and position $q_n$, 
described by the Hamiltonian
\begin{equation}
{H} = \sum_{n=1}^N \left[{p_n^2\over 2m} + U(q_n) +
V(q_{n+1}-q_{n})\right] \quad .
\label{optical}
\end{equation}
The potential $V(x)$ accounts for the nearest-neighbour interactions between consecutive particles,
while the on-site potential $U(q_n)$ takes into account the possible interaction with an
external environment (either a substrate, or some three-dimensional matrix).
The corresponding evolution equations are
\begin{equation} 
m{\ddot q}_n = -U'(q_n) - F(r_{n}) + F(r_{n-1})  \quad , \quad n=1,\ldots,N \, ,
\label{eqmot} 
\end{equation}  
where $r_n=q_{n+1}-q_n$, $F(x)=- V'(x)$, and the prime denotes a derivative with respect to the argument.
Usually $q_n$ denotes the longitudinal position along the chain, so that 
\begin{equation}
L= \sum_{n=1}^N r_n  \; ,
\label{acoustic2}
\end{equation}
represents the total length of the chain (which, in the case of fixed b.c., is a constant of motion).
Different kinds of boundary conditions may and will be indeed used in the various cases.
For instance, if the particles are confined in a simulation ``box"
of length $L$ with periodic boundary conditions, 
\begin{equation}
q_{n+N} \;=\; q_{n} \,+\, L  \quad .
\label{pbc}
\end{equation}
Alternatively one can adopt a lattice interpretation,
in which case, the (discrete) position is $z_n = an$ (where $a$ is lattice spacing), while $q_n$
is a transversal displacement. Thus, the chain length is obviously equal to $Na$.

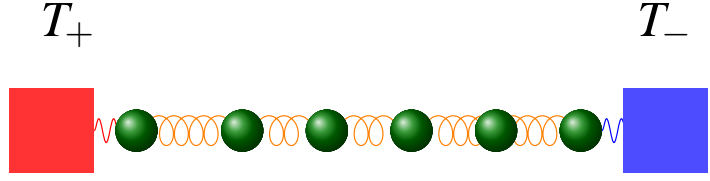
\begin{figure}[t]
\begin{center}

\resizebox{0.8\textwidth}{!}{%
 \begin{tikzpicture}
\fill[red!80!white] (-4,-3) rectangle (-2,-1) ;
\draw[thick,red,decorate,decoration={snake,aspect=0.5,amplitude=8}] (-2.,-2) -- (-1,-2);
\draw[thick,orange,decorate,decoration={coil,aspect=0.7,amplitude=10}] (-.8,-2) -- (1.5,-2);
\draw[thick,orange,decorate,decoration={coil,aspect=0.7,amplitude=10}] (1.8,-2) -- (3.5,-2);
\draw[thick,orange,decorate,decoration={coil,aspect=0.7,amplitude=10}] (3.8,-2) -- (5.5,-2);
\draw[thick,orange,decorate,decoration={coil,aspect=0.7,amplitude=10}] (5.8,-2) -- (9.5,-2);
\fill[ball color= green!50!black] (-1.,-2) circle (.5);
\fill[ball color= green!50!black] (1.5,-2) circle (.5);
\fill[ball color= green!50!black] (3.5,-2) circle (.5);
\fill[ball color= green!50!black] (5.5,-2) circle (.5);
\fill[ball color= green!50!black] (7.5,-2) circle (.5);
\fill[ball color= green!50!black] (9.5,-2) circle (.5);\draw[thick,blue,decorate,decoration={snake,aspect=0.5,amplitude=8}] (10,-2) -- (11.5,-2);
\fill[blue!70!white] (10.5,-3) rectangle (12.5,-1);
\draw (-2.6,0.5) node[scale=4]{$T_+$};
\draw (11.5,0.5) node[scale=4]{$T_-$};
\end{tikzpicture}
}%
\end{center}
\caption{A one-dimensional chain of coupled oscillators
interacting with two thermal reservoirs ad different temperatures $T_+$ and $T_-$.}
\label{fig:baths} 
\end{figure}

The Hamiltonian (\ref{optical}) is generally a constant of motion.
In the absence of an on-site potential ($U=0$), the total momentum is conserved, as well,
\begin{equation}
P= \sum_{n=1}^N p_n \equiv \sum_{N=1} m \dot q_n\; .
\label{acoustic}
\end{equation}
Since we are interested in heat transport, one can set $P = 0$ 
(i.e., we assume to work in the center--of--mass reference frame) without
loss of generality. 
As a result, the relevant state variables of microcanonical equilibrium are the 
specific energy (i.e., the energy per particle) $e=H/N$ and the 
elongation $\ell=L/N$ (i.e., the inverse of the particle density).
On a microscopic level, one can introduce three local densities,
namely $r_n$, $p_n$ and 
\begin{equation}
    e_n =  {p_n^2\over 2m_n} + \frac12 \bigg[ V(r_n) +
    V(r_{n-1}) \bigg] \quad ,
    \label{loce}
\end{equation}
which, in turn, define a set of currents through three (discrete) continuity equations. 
For instance,  the energy current is defined as  
\begin{eqnarray}
&&\dot e_n = j_{n-1} - j_n \\
&&j_n = \frac{1}{2} a (\dot q_{n+1}+\dot q_n ) \,  F(r_n) \quad .
\label{hf2}
\end{eqnarray}
The definition (\ref{hf2}) is related to the general
expression of the energy current, originally derived by Irving and Kirkwood that is valid 
for every state of  matter (see e.g. \cite{Kubo1991}) that, in one dimension, reads
\begin{equation} 
j_n = \frac{1}{2} (q_{n+1} - q_n) ({\dot q}_{n+1} + {\dot q}_n) \,
     F(r_n) + {\dot q}_n e_n \, .
\label{jsolid} 
\end{equation}
In the case of lattice systems, where we assume 
the limit of small oscillations (compared to the lattice spacing) or 
in the lattice field interpretation, one can recover formula (\ref{hf2})
setting $ q_{n+1} - q_n= a$ in the first term and
neglecting the second one \cite{LLP03}.
The expression (\ref{jsolid}) is useful
in the opposite limit of freely colliding particles, where the only relevant interaction
is the repulsive part of the potential, that is responsible for elastic 
collisions. There, the only contribution to the flux arises from the 
kinetic term of $e_n$, i.e.
\begin{equation} 
   j_n  \;\approx\; \frac{1}{2} m_n{\dot q}_n^3 \quad .
\label{lj612}
\end{equation} 

Having set the basic definitions, let us now introduce some specific models.
A first relevant example is the harmonic chain, where the potential $V$ is quadratic
(and $U=0$).  From the point of view of transport properties, 
we expect this system to behave like a ballistic conductor. 
The heat flux decomposes into the sum of independent contributions associated 
to the various eigenmodes. This notwithstanding, this model proves useful, as it
allows addressing general questions about the nature of stationary nonequilibrium states.
This includes the role of disorder (either in the masses or the spring constants), of
boundary conditions, and quantum effects. Since the linear case (classical and quantum)
will be treated in detail in 
Chapter 2, here we focus on the anharmonic problem.
In this context, the most paradigmatic example 
is the Fermi--Pasta--Ulam (FPU) model 
\cite{Payton67,Nakazawa1970,Kaburaki93}
\begin{equation}
V(r_n)\;=\; \frac{k_2}{2}\,(r_n-a)^2+ \frac{k_3}{3}\,(r_n-a)^3 + 
\frac{k_4}{4}\, (r_n-a)^4 \quad .
\label{fpu}
\end{equation} 
Following the notation of the original work \cite{Fermi1955}, the couplings 
$k_3$ and $k_4$ are denoted by $\alpha$ and $\beta$ respectively and 
historically this model with is sometimes referred to as the ``FPU-$\alpha\beta$'' model.
Also, the quadratic plus quartic ($k_3=0$) potential is termed the ``FPU-$\beta$'' model. 
Notice that upon introducing the displacement $u_n = q_n - na$ from the equilibrium position,
$r_n$ can be rewritten as $u_{n+1}-u_n+a$, so that the lattice spacing $a$ disappears from the equations.

Another interesting model is the Hard Point Gas (HPG), where the interaction potential is
\cite{Casati1986,H99,Grassberger02}
\begin{displaymath}
 V(y)= 
\begin{cases}
     \infty & \quad \mbox{$y=0$}\\
     0 & \quad \mbox{otherwise}
\end{cases} \; .
\label{uhpg}     
\end{displaymath}
The dynamics consist of successive collisions between neighbouring particles,
\begin{equation}
v_n'=\frac{m_n-m_{n+1}}{m_n+m_{n+1}} v_n+ \frac{2 m_{n+1}}{m_n+m_{n+1}}
v_{n+1} \quad , \quad
v_{n+1}'= \frac{2 m_{n}}{m_n+m_{n+1}}
v_n-\frac{m_n-m_{n+1}}{m_n+m_{n+1}} v_{n+1} \quad , 
\label{coll}
\end{equation}
where $m_n$ is the mass of the $n$th particle, $v_n= \dot q_n$ and the primed variables 
denote the values after the collision. For equal masses the model is completely integrable,
as the set of initial velocities is conserved during the evolution. In order to 
avoid this peculiar situation, it is customary to choose alternating values, such as
$m_{n}=m$ ($rm$) for even (odd) $n$. This type of dynamical systems are particularly appropriate
for numerical computation as they do not require the numerical integration of nonlinear differential
equations. In fact, it is sufficient to determine the successive collision times 
and update the velocities according to Eqs.~(\ref{coll}). The
only errors are those due to machine round-off. Moreover, the simulation can be
made very efficient by resorting to fast updating algorithms. In fact, since the
collision times depend only on the position and velocities of neighbouring particles,
they can be arranged in a heap structure and thereby simulate the dynamics with an
an event driven algorithm~\cite{Grassberger02}.

Another much studied model involves the Lennard--Jones potential, 
that in our units reads \cite{Mareschal88,Lepri2005}
\begin{equation}
V(r) \;=\;  {1\over 12}
\bigg({1 \over r^{12}}\, -\, {2\over  r^6}\,+\, 1 \bigg)\quad .
\label{lj}
\end{equation}
For computational purposes, the coupling parameters have been fixed in such a
way as to yield the simplest form for the force. With this choice, $V$ has a
minimum in $y=1$
and the resulting dissociation energy is $V_0=1/12$. For the sake of
convenience, the zero of the potential energy is set in $y=1$. In one-dimension, 
the repulsive term ensures that the ordering is preserved
(the particles do not cross each other).

In the presence of a substrate potential $U$, the invariance $q_l \to q_l+const.$ is
broken and the total momentum $P$ is no longer a constant of motion.
Accordingly, all branches of the dispersion relation have a gap at zero wavenumber. We 
therefore refer to them as {\it optical} modes. 
An important subclass is the one in which $V$ is quadratic, which can be regarded as a discretization
of the Klein-Gordon field: relevant examples are the Frenkel-Kontorova \cite{Gillan85,Hu1998}
and ``$\phi^4$'' models \cite{Aoki00} which, in suitable units, correspond to 
$U(y) = 1-\cos(y)$ and $U(y) = y^2/2+y^4/4$, respectively.              
Another toy model that has been studied in some detail 
is the ding-a-ling system~\cite{Casati84}, where $U$ is quadratic and the nearest-neighbor
interactions are replaced by elastic collisions.

We will always deal with genuine nonintegrable dynamics. For the FPU model this means working 
with high enough energies/temperatures to avoid all 
the difficulties induced by quasi-integrability and the 
associated slow relaxation to equilibrium. For the 
diatomic HPG this requires fixing a mass-ratio $r$ not too close to unity.

\section{Signatures of anomalous transport}
\label{sec:signatures}

The results emerged from a long series of works can be summarized
as follows. Models of the form (\ref{eqmot}) with $U(q)=0$ typically display  
\textit{anomalous} transport and relaxation features, this meaning that
(at least) one of the following phenomena has been reported:
\begin{itemize}

\item The finite-size heat conductivity $\kappa(L)$  diverges  in
the limit of a large system size $L\to \infty$ \cite{LLP97} as
\footnote{ For historical reasons
two of the scaling exponents introduced in
this subsection are conventionally denoted by the same  Greek letters,
$\alpha$ and $\beta$,  adopted for the FPU models
described in Section 2.} 
\[
\kappa(L) \;\propto\; L^\alpha
\]
This means that
this transport coefficient is ill-defined in the thermodynamic limit;

\item The equilibrium correlator of the
energy current displays a nonintegrable power-law decay, 
\begin{equation}
\langle J(t)J(0)\rangle \propto t^{-(1-\delta)}
\label{longtail}
\end{equation}
with $0\le\delta < 1$,  
for long times $t\to \infty$ \cite{Lepri98a}. Accordingly, 
the Green-Kubo formula yields an infinite value of the conductivity; 

\item Energy perturbations propagate superdiffusively \cite{Denisov03,Cipriani05}: 
a local perturbation of the energy broadens and its variance 
$\sigma^2$ 
grows in time as  
\begin{equation}
\sigma^2(t)  \propto t^{\beta}
\end{equation}
with $\beta > 1$;
\item Relaxation of spontaneous fluctuations is fast 
(i.e. superexponential) \cite{Lepri2005}: 
at variance with standard hydrodynamics, the typical
decay rate in time of fluctuations at wavenumber $k$, $\tau(k)$, is found
to scale as 
$$\tau(k) \sim |k|^{-z}$$ 
(with $z<2$).

\item Temperature profiles in the nonequilibrium steady states 
are nonlinear, even for vanishing applied temperature 
gradients.

\end{itemize}

Altogether, these features can be summarized by saying that the usual 
Fourier's law \textit{does not hold}: the kinetics of energy carriers is so
correlated that they are able to propagate \textit{faster} than in the
the standard (diffusive) case. 

Numerical studies ~\cite{LLP03} 
indicate that anomalies occur generically in 1 and 2D, whenever
the conservation of energy, momentum and length holds.
This is related to the existence of
long-wavelength (Goldstone) modes  (an acoustic phonon branch in the linear
spectrum of (\ref{eqmot}) with $U=0$) that are very weakly damped.
Indeed, it is sufficient to add external (e.g. substrate) forces, to make
the anomalies disappear.  

Let us now discuss these features in more detail.

\subsection{Diverging finite-size conductivity}

A natural way to simulate a heat conduction experiment consists in putting the
system in contact with two heat reservoirs operating at different temperatures
$T_+$ and $T_-$ (see Fig. \ref{fig:baths}). This requires a suitable modeling of interaction with 
the enviroment. Several methods, based on both deterministic
and stochastic algorithms, have been proposed. A more detailed presentation 
can be found in  \cite{LLP03}
and \cite{DHARREV}. A simple and widely used choice consists in adding Langevin-type forces
on some chain subsets. If this is done on the first and the last site of a finite chain 
($n=1,\ldots,N$), it is obtained
\begin{equation}
  \ddot q_n =  - F_n + F_{n-1} +
 \delta_{n1}(\xi_+ - \lambda \dot q_1) + \delta_{nN}(\xi_- -\lambda \dot q_N) \quad ,
\label{eq:homlin1}
\end{equation}   
where we assume unitary--mass particles, while $\xi_\pm$'s are two independent 
Gaussian processess with zero mean and variance $2\lambda  k_BT_\pm$
($k_B$ is the Boltzmann constant). The coefficient $\lambda$ is
the coupling strength with the heat baths.

After a long enought transient, an off-equilibrium stationary state sets in, with a
net heat current flowing through the lattice. 
\footnote{From the mathematical point of view, the existence of 
a unique stationary measure is a relevant question and 
has been proven in some specific cases models of this class, see the review \cite{BLRB00} and \cite{EPRB99,Eckmann2000}.}
The thermal conductivity $\kappa$ of
the chain is then estimated as the ratio between the time--averaged flux
$\overline j$ and the overall temperature gradient $(T_+-T_-)/L$, where 
$L$ is the chain length. Notice that,
by this latter choice, $\kappa$ amounts to an effective transport coefficient,
including both boundary and bulk scattering mechanisms. The average
$\overline j$ can be estimated in several equivalent ways, depending on the
employed thermostatting scheme. One possibility is to directly measure the
energy exchanges with the two heat reservoirs \cite{LLP03,DHARREV}. 
A more general  (thermostat-independent) definition consists in averaging the 
heat flux as defined by (\ref{jsolid}).

\begin{figure}[t]
\begin{center}
\includegraphics[width=7.5cm]{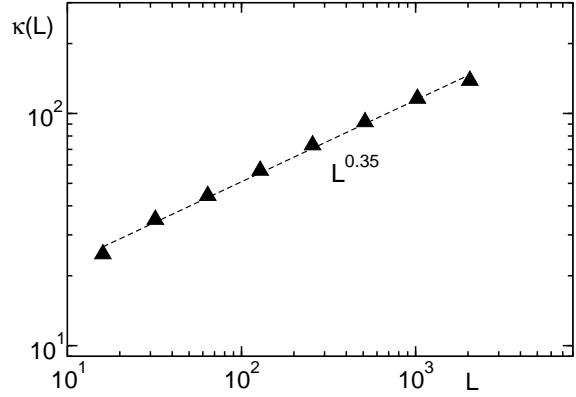}
%
%
\caption{Scaling of the finite-size conductivity for the FPU-$\alpha\beta$  model:
with energy $e=1$ and cubic coupling constant $\alpha=0.1$.}
\end{center}
\label{fig:conduc}       
\end{figure}

As a result of many independent simulations performed with the above-described
methods, it is now established that $\kappa \propto L^\alpha$ for 
$L$ large enough. Fig.\ref{fig:conduc} illustrates the typical outcome of simulations 
for the FPU chain. 

\subsection{Long-time tails}

In the spirit of linear--response theory, transport coefficients can be
computed from equilibrium fluctuations of the associated currents. 
More precisely, by introducing the total heat flux
\begin{equation}
J = \sum_n j_n \quad ,
\label{tot-heat}
\end{equation}
the Green-Kubo formalism tells us that heat conductivity is given by 
the expression
\begin{equation}
\kappa \;=\; \frac{1}{k_BT^2}\lim_{t\to\infty} \lim_{N\to\infty} 
\frac{1}{N}\int_0^t 
 \, dt' \, \langle J(t')J(0) \rangle  \quad ,
\label{GK}
\end{equation}
where the average is performed in a suitable equilibrium ensemble, e.g.
microcanonical with zero total momentum ($P=0$). 

A condition for the formula (\ref{GK}) to give a well--defined heat conductivity
is that the time integral is convergent. This is clearly not the case when the
current correlator vanishes as in (\ref{longtail})
 with $0\le \delta < 1$. Here, the integral diverges as
$t^\delta$ and we may thus define a finite--size conductivity $\kappa(L)$ by
truncating the time integral in the above equation to $t\approx L/c$, where $c$
is the sound velocity. Consistency with the definition of the power--law 
divergence of $\kappa(L)$ implies
$\alpha=\delta$. The available data agrees with this expectation, thus
providing an independent method for estimating the exponent $\alpha$.

\begin{figure}[t]
\begin{center}
\includegraphics[width=7.5cm]{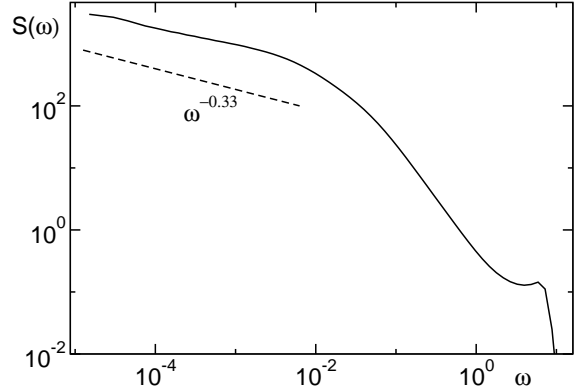}
%
%
\caption{Spectrum of energy current for the
FPU-$\alpha \beta$ model, same parameters as in previous figure.}
\end{center}
\label{fig:cspectrum}       
\end{figure}

For later purposes, we mention that, by means of the Wiener--Khintchine
theorem, one can equivalently extract $\delta$ from the low-frequency 
behaviour of the spectrum of current fluctuations 
\begin{equation}
S(\omega) \equiv \int \, d\omega \langle J(t)J(0)\rangle {\rm e}^{i\omega t}
\label{Fourier}
\end{equation}
that displays a low--frequency singularity of the form $S(\omega) \propto
\omega^{-\delta}$ (see Fig.~\ref{fig:cspectrum}).
 From the practical point of view, this turns out to be the
most accurate numerical strategy, as divergencies are better estimated than 
convergences to zero.

\subsection{Diffusion of perturbations}

Consider an infinite system at equilibrium with
a specific energy $e_0$ per particle and a total momentum $P = 0$.
Let us  perturb it by increasing 
the energy of a subset of adjacent particles
by some preassigned amount $\Delta e$ and denote with
$e(x,t)$ the energy profile evolving from such a perturbed 
initial condition (for simplicity, we identify $x$ with the average particle 
location $n\ell$). We then ask how the perturbation 
\begin{equation}
\delta e(x,t) = \langle e(x,t) - e_0 \rangle
\label{fdef}
\end{equation}
behaves in time and space \cite{Helfand60}, where the angular brackets 
denote an ensemble average 
over independent trajectories. Because of energy conservation,
$\sum_n \delta e(n\ell,t) = \Delta e$ remains constant 
at any time: $\delta e(x,t)$ 
can be interpreted as a probability density (provided it is also
positive-defined and normalized). 

\begin{figure}[t]
\begin{center}
\includegraphics[width=7.5cm]{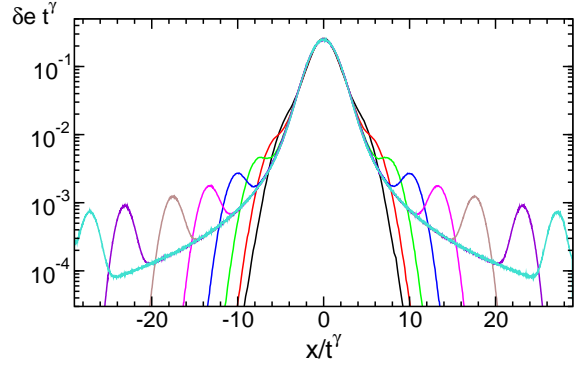}
%
%
\caption{Spreading of infinitesimal perturbations in the HPG model:
rescaled perturbation profiles at different times 
$t= 40, 80, 160, 320, 640, 1280, 2560, 3840$ (the width increases with time), 
with $\gamma=3/5$. }
\end{center}
\label{fig:spreading}       
\end{figure}

For sufficiently long time $t$ and large $x$, 
one expects $\delta e(x,t)$ to scale as
\begin{equation}
\delta e(x,t)=t^{-\gamma}\mathcal G (x/t^{\gamma}) 
\label{eq:scail}
\end{equation}
for some probability distribution $\mathcal G$ and a scaling parameter 
$0\le\gamma\le 1$.
The case $\gamma=1/2$ corresponds to a normal
diffusion and to a normal conductivity. On the other
hand, $\gamma=1$ corresponds to a ballistic motion and to a linear
divergence of the conductivity.
Consequently, a $\gamma$-value larger than $1/2$ implies a
superdiffusive behaviour of the macroscopic evolution of the energy
perturbation \cite{Denisov03}. 
In Fig.~\ref{fig:spreading}, the evolution of infinitesimal energy
perturbations is reported in the case of the HPG~\cite{Cipriani05}:
a very good data-collapse is reported for $\gamma=3/5$.

Remarkably, the above results can be rationalized in terms of a very 
simple random dynamics: the \textit{L\'evy walk model} \cite{Blumen1989,Zaburdaev2015}. 
Consider a point particle that moves
ballistically in between successive ``collisions'', whose time
separation is distributed according to a power law, $\psi(t) \propto
t^{-\mu -1}$, $\mu >0$, while its velocity is chosen from a symmetric
distribution $\Psi(v)$. By assuming a $\delta$-like distribution, 
$\Psi(v) =(\delta(v-\tilde v) + \delta(v+\tilde v))/2$, the propagator
$P(x,t)$ (the probability distribution function to find in $x$ at time $t$, a
particle initially localized at $x=0$) can be written as
$P(x,t) = P_{L}(x,t) + t^{1-\mu}[\delta(x-\tilde vt) + \delta(x-\tilde vt)]$
where \cite{Blumen1989} 
\begin{equation}
   P_L(x,t) \propto    
   \left \{      
   \begin{array}{ll}
          t^{-1/\mu}\, \exp \left[-(\eta x/t^{1/\mu})^2\right]
                                        & |x| < t^{1/\mu} \\
          t \, x^{-\mu-1}  & t^{1/\mu} , |x| <  \tilde v t\\
          0   &|x| > \tilde v t
    \end{array}
    \right. \quad,
    \label{LW}
\end{equation}
where $\eta$ is a generalized diffusion coefficient.
From the evolution of the perturbation profile, it is possible to infer the
exponent $\alpha$ of the thermal conductivity.
In fact, in \cite{Denisov03}  it has been argued that the exponents $\alpha$,
$\beta$  (the growth rate of the mean square displacement, 
$\sigma^2(t)= \sum_{n}\, n^{2}\delta e(x=n\ell,t) \,\propto\, t^{\beta}$)  
and $\gamma = 1/\mu$ are linked by the following relationships,
\begin{equation}
\alpha\; =\; \beta\;-\;1\;=\; 2\;-\;\frac{1}{\gamma} 
\label{alfa2}.
\end{equation}
In particular, we see that the case $\gamma=1/2$ corresponds to normal
diffusion ($\beta=1$) and to a normal conductivity ($\alpha=0$). On the other
hand, $\gamma=1$ corresponds to a ballistic motion ($\beta=2$) and to a linear
divergence of the conductivity ($\alpha =1$). The numerically observed value
$\gamma=3/5$ corresponds to an anomalous divergence with $\alpha=1/3$.

The spreading of the wings can be accounted by means of a model which
allows for velocity fluctuations \cite{Zaburdaev2011,Denisov2012}, which originates
from wave dispersion. Assigning
smoother velocity distributions $\Psi(v)$ leads to 
broadening of $\delta$ side-peaks, but does not affect the shape and the scaling behavior of
the bulk contribution $P_L(x,t)$, which scales, as predicted in 
Eq.~(\ref{eq:scail}), with the exponent $\gamma=1/\mu$.

An alternative way to study finite amplitude perturbations 
is by looking directly at the behavior of the nonequilibrium 
correlation function of the energy density \cite{Zhao06},
\begin{equation}
C_e(x,t) = \langle \delta e(y,\tau)\delta e (x+y,t+\tau)\rangle \quad ,
\end{equation}
where the angular brackets denote a spatial as well as a temporal average
over the variables $y$ and $\tau$, respectively.
At $t=0$,  $C_e(x,0)$ is a $\delta$ function in
space. Moreover, in the microcanonical ensemble, energy conservation implies 
that the area $\int dx C_e(x,t)$ is constant at any time. By 
assuming that $C_e(x,t)$ is normalized to a unit area, its behaviour is 
formally equivalent to that of a
diffusing probability distribution. This allows one
to determine the scaling behavior of the heat conductivity from the
growth rate of the variance of $C_e(x,t)$ \cite{Zhao06}. As the determination
of the variance is troubled by the fluctuating tails, it is preferable to proceed 
by looking at the decay of the maximum $C_e(0,t)$, that is statistically more reliable. 
An interesting relation between correlation function and anomalous heat transport 
has been pointed out recently \cite{Liu2014} and is reviewed in Chapter 6.

\subsection{Relaxation of spontaneous fluctuations}

The above discussion suggests that scaling concepts can be of great
importance in dealing with thermal fluctuations of conserved quantities.
The evolution of a fluctuation of wavenumber $k$ excited at $t=0$ is 
described by its correlation function.
For 1D models like (\ref{optical}) one of such functions is defined by considering 
the relative displacements $u_n=q_n-n\ell$ and defining the
collective coordinates through the discrete transform
\begin{equation}
U(k,t)  \;=\; {1\over N}\,\sum_{n=1}^N \, u_n \exp(-ik n) \quad .
\label{dens}
\end{equation}
By virtue of the periodic boundaries, the allowed
values of the  wavenumbers $k$ are integer multiples of $2\pi/N$.
We then define the dynamical structure factor, namely the square modulus of 
the temporal Fourier transform of the particle displacements as
\begin{equation}
S(k,\omega) \;=\; 
\big\langle \big| U(k,\omega )\big|^2 \big\rangle  \quad .
\label{strutf}
\end{equation}
The angular brackets denote an average over an equilibrium ensemble.

\begin{figure}[t]
\begin{center}
\includegraphics[width=7.5cm,clip=true]{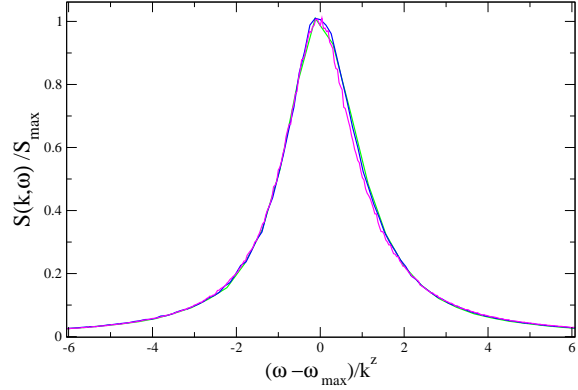}
\caption{FPU$\alpha\beta$  model: Check of dynamical scaling for the 
dynamical structure factors $\alpha=0.1$, $N=4096$, $e=0.5$ and 
four different wavenumbers  $k=2,4,8,16$ (in units of $2\pi/N$). The best estimate of
the dynamical exponent is $z=1.5$. 
}
\end{center}
\label{fig:4}       
\end{figure}

For sufficiently small wavenumbers $k$, the dynamical structure factor $S(k,\omega)$
usually displays sharp peaks at finite frequency, whose position is proportional 
to the wavenumber $\omega_{max} = c|k|$; $c$ is naturally interpreted as the 
phonon sound speed. The data in Fig. \ref{fig:4} show that long-wavelength 
correlations, $k \to 0$, obey 
\textit{dynamical scaling}, i.e. there exist a function $f$ such that 
\begin{equation}
S(k,\omega) \sim  f\left( \frac{\omega-\omega_{max}}{k^{z}} \right).
\label{dynscal} 
\end{equation}
for $\omega$ close enough to $\omega_{max}$.
The associated linewidths are a measure of the fluctuation's
inverse lifetime. Simulations indicate that these
lifetimes scale as $k^{-z}$ with $z\approx 1.5$. Thus the 
behavior is different from the diffusive one where one would 
expect $z=2$. As explained above, one may
think of this as a further signature of an underlying superdiffusive process, 
intermediate between standard Brownian motion and ballistic propagation. 

Other correlation functions can be defined similarly and obey some
form of dynamical scaling. For instance, one could consider the structure factor 
$S_e(k,\omega)$ associated with the local energy density $e_n$, defined in (\ref{loce}). 
It has a large central component (as a result of the heat modes) and a ballistic one
(following from the sound modes). If we assume that the low-frequency part is dominated by 
the heat-mode scaling, we should have for $\omega\to 0$ 
\begin{equation}
S_e(k,\omega) \sim  g( \omega/q^{5/3} ) \; ,
\label{dynscale} 
\end{equation}
with $g$ being a suitable scaling function.

The origin of the nontrivial dynamical exponents are to be traced back to the 
nonlinear interaction of long-wavelength fluctuations. For a chain of coupled 
anharmonic oscillators with three conserved fields ($H$, $L$, and $P$), a linear theory would 
yield two propagating sound modes and one diffusing heat mode, all of the 
three diffusively broadened.
In contrast, the nonlinear theory predicts that, at long times, the sound mode correlations 
satisfy Kardar-Parisi-Zhang scaling, while the heat mode correlations follow a L\'evy-walk scaling. 
Various spatiotemporal correlation functions of Fermi-Pasta-Ulam chains
and a comparison with the theoretical predictions can be found in \cite{Das2014a}.

\subsection{Temperature profiles}

Anomalous transport manifests itself also in the shape of the steady-state 
temperature profiles. For chains in contact with two baths like 
in Eq.~(\ref{eq:homlin1}), one typically observes that the kinetic 
temperature profile $T_n=\langle p_n^2\rangle$ is distinctly 
nonlinear also for small temperature differences $\Delta T$.
For fixed $\Delta T$, the profile typically satisfies a ``macroscopic" scaling, 
$T_n=T(n/L)$ for $L\to \infty$ with $T(0)=T_+$ and $T(1)=T_-$.
\footnote{Temperature discontinuities may appear at the chain boundaries.
This is a manifestation
of the well-known Kapitza resistance, the temperature discontinuity
arising when a heat flux is maintained across
an interface among two substances. This discontinuity is the result of a
boundary resistance, that is explained as a ``phonon mismatch" 
between the two media: see \cite{Aoki01} for a discussion of the class of 
models at hand. 
}

In view of the above correspondence with L\`evy processes it may be 
argued that this feature too could be described in terms of 
anomalous diffusing particles in a finite domain and subject to 
external sources that steadily inject particles through its boundaries.
The idea is to interpret the local temperature $T(x)$ as the density $P(x)$
of suitable random walkers.
A general stochastic model can be defined as follows \cite{Lepri2011}.
Let $n$ denote the position of a discrete-time  
random walker on a finite one-dimensional lattice ($1\le n\le N)$. 
In between consecutive scattering events, the particle either jumps 
instantaneously (L\'evy flight - LF) or moves with unit velocity 
(L\'evy walk - LW) over a distance of $m$ sites, 
that is randomly selected according to the step-length distribution 
\begin{equation}
\label{eq:dist}
\lambda_m \; = \; \frac{q}{|m|^{1+\mu}},\quad \lambda_0=0 \; ,
\end{equation}  
which is the discrete analogous of the $\psi$ distribution defined above, 
with $\mu$  ($1\leq \mu \le 2$) being the L\'evy exponent and $q$ a normalization constant.
The process can be formulated by introducing the vector ${\bf W}\equiv\{ W_n(t)\}$,
where $W_n$ is the probability for the walker to undergo a scattering event at
site $n$ and time $t$. It satisfies a master equation, which, for LFs, writes
\begin{equation}
{\bf W}(t+1) = {\bf Q}{\bf W}(t) +{\bf S} ,
\label{eq:sme}
\end{equation}
where ${\bf S}$ accounts for the particles steadily injected from external
resorvoirs; ${\bf Q}$ is a
matrix describing the probability of paths connecting pair of sites. In the
simple case of absorbing BC, it is readily seen that $Q_{ji}$ is equal to the
probability $\lambda_{j-i}$ of a direct flight, as from Eq.~(\ref{eq:dist}). 
In the LW case, the ${\bf W}$ components in the r.h.s.
must be estimated at different times (depending on the length of the path
followed from $j$ to $i$) \cite{Klafter1987}. Since, the stationary solution is
the same in both cases, this difference is immaterial, and is easier to refer to
LFs, since Eq.~(\ref{eq:sme}) can be solved iteratively. Note that
in the LF case, $W_i$ is equal to the density $P_i$ of particles at site $i$,
while for the LW, $P_i$ includes those particles that 
are transiting at the $i$th site during a ballistic step. 

The source term is fixed by assuming that the reservoir is a semi-infinite
lattice, homogeneously filled by L\'evy walkers of the same type as those
residing in the domain. This amounts to defining $S_m = s\, m^{-\mu}$, where
$s$ measures the density of particles and $m$ the distance from the
reservoir. It is easy to verify that in the presence of two identical 
reservoirs at the lattice ends, the density is constant
(for any $N$), showing that our definition satisfies a kind of
``zeroth principle",  as it should.

\begin{figure}[t]
\begin{center}
\includegraphics[width=7.5cm]{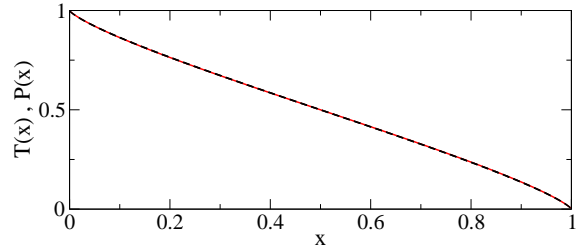}
\caption{Temperature profile $T(x)$ of the oscillator chain with conservative noise 
with free boundary condition and $\lambda=\gamma=1$
(solid line) and density profile $P(x)$ for the master equation
with reflection coefficient $r=-0.1$ (dashed line).
}
\end{center}
\label{fig:profile}       
\end{figure}

In the nonequilibrium case, it is not necessary to deal with two reservoirs. The
linearity of the problem teaches us that it is sufficient to study the case of a
single reservoir, that we assume to be in $n=0$: the effect of, say, a second
one on the opposite side can be accounted for by a suitable linear combination.
For large-enough $N$ values, the steady-state density 
depends on $n$ and $N$ through the combined variable $x=n/N$, i.e. $P(x)=P_n$. 
As seen in Fig.~\ref{fig:profile}, $P$ vanishes for $x \to 1$ because on that side 
the absorbing boundary is not accompanied by an incoming flux of particles.

Altogether, upon identifying the particle density with the temperature, 
the profile can be viewed as a stationary solution of the stationary 
Fractional Diffusion Equation (FDE)
\begin{equation}
D^\mu_x P=-\sigma(x)
\label{fracdiff}
\end{equation}
on the interval $0\le x \le 1$ (see e.g. Ref.~\cite{Zoia07} 
and references therein for the 
definition of the integral operator $D^\mu_x$). The source term $\sigma(x)$ 
must be chosen so as to describe the effect of the external reservoirs.  
A condition to be fulfilled is that two identical reservoirs yield
a homogeneous state $T(x)=const.$\,.  
Using the integral definition of $D^\mu_x$ \cite{Zoia07}, it can be 
shown that this happens for $\sigma(x) = \sigma_{eq}(x) 
\equiv x^{-\mu}+(1-x)^{-\mu}$ 
(we, henceforth, ignore irrelevant proportionality constants). 
It is thus natural to associate $\sigma(x) = x^{-\mu}$ to the 
nonequilibrium case with a single source in $x=0$.
The numerical solution of the FDE agrees 
perfectly with the stationary solution of the discrete model, thus
showing that long-ranged sources are needed to reproduce the 
profiles in the continuum limit.

A distinctive feature of the profile 
is that it is not analytic at the boundaries. Indeed, the data for 
$x\to 0$ are well fitted by
\begin{equation}
P(x) \;=\; P(0) + Cx^{\mu_m}
\end{equation}
(the same behavior occurs for $x\to 1$, as the profiles are
symmetric). In view of the similarity with the shape of the liquid
surface close to a wall, we metaphorically term $\mu_m$ as the 
\textit{meniscus exponent}. Such 
nonanalytic behavior is peculiar of anomalous kinetics, 
as opposed to the familiar linear shape in standard diffusion.   
For the above discussed case of absorbing BC, we find that
$\mu_m\approx \mu/2$. This value is consistent with the  
singular behavior of the eigenfunctions of $D^\mu_x$
\cite{Zoia07}. In the general case, by assuming a linear dependence
of $\mu_m$ on  both $r$, and $\mu$,  it has been conjectured that~\cite{Lepri2011}
\begin{equation}
\label{eq:conj}
\mu_m \;=\;\frac{\mu}{2} + r \left(\frac{\mu}{2}-1\right)\,.
\end{equation}
This expression is consistent with the $\mu_m=\alpha/2$ value found 
above for $r=0$. Moreover, for $\alpha=2$ (normal diffusion) 
it yields $\mu_m=1$, as it should.

Let us now compare this probability distribution of the above 
process with the temperature profiles in one-dimensional systems displaying 
anomalous energy transport. It is convenient to refer to
a chain of harmonic oscillators coupled with two Langevin heat 
baths (with a damping constant $\lambda$), and with random collisions 
that exchange the velocities of neighboring particles with a rate $\gamma$ \cite{DLLP08}.
On the one hand, this model has the advantage of allowing for an exact solution of the 
associated Fokker-Planck equation \cite{Lepri2009}; on the other hand it
is closely related to a model that has been proved to display a L\'evy-type
dynamics \cite{BBO06}.

In Fig.~\ref{fig:profile} we compare the temperature profile $T(x)$ (suitably shifted
and rescaled) of the heat-conduction model \cite{Lepri2009} with free BC and the solution of our
discrete L\'evy model with a reflection coefficient $r=-0.1$.
Since they are
essentially indistinguishable, we can conclude that the L\'evy interpretation does
not only allow explaining the anomalous scaling of heat conductivity
\cite{Cipriani05}, but also the peculiar shape of $T(x)$. The weird (negative)
value of $r$ can be justified a posteriori by introducing two families of 
walkers  and interpreting the reflection 
as a change of family. The relevant quantity to look at is the difference between the
densities of the two different families. The reason why it is necessary to invoke the
presence of such two families and their physical meaning in the context
of heat conductivity is an open problem.

In the case of a chain with fixed BC, the temperature profile $T(x)$ 
can be computed analytically \cite{Lepri2009} and it is thereby found 
that $\mu_m=1/2$. By inserting this value in 
Eq.~(\ref{eq:conj}) and recalling that $\mu=3/2$, we find that $r=1$, 
i.e. the fixed-BC $T(x)$ corresponds to the case of perfectly reflecting 
barriers. Unfortunately, this (physically reasonable) result could not
be tested quantitatively. Indeed, it turns out that finite-size
corrections become increasingly important upon increasing $r$, and for 
$r$ close to 1, it is practically impossible to achieve convergence
to the steady-state. 

The description of the steady state in terms of L\'evy walk has been further 
investigated in \cite{Dhar2013}. 
The authors calculate exactly the average heat current, the large deviation function of its fluctuations, 
and the temperature profile in the steady state. 
The current is nonlocally connected to the temperature gradient. Also, all the cumulants of the current 
fluctuations have the same system-size dependence in the open geometry
as those of deterministic models like the HPG. The authors investigated also 
the case of a ring geometry and argued that a size-dependent cutoff time 
is necessary for the L\'evy-walk model to behave like in the deterministic case. 
This modification does not affect the results on transport in the open geometry 
for large enough system sizes.

\section{Universality and theoretical approaches}
\label{sec:universality}

In view of their common physical origin, it is
expected that the exponents describing the different processes will be related to each
other by some ``hyperscaling relations''. 
Their value should be
ultimately dictated by the dynamical scaling of the underlying dynamics.
Moreover, one can hope that they are largely independent of the
microscopic details, thus allowing for a classification  of anomalous behavior
in terms of ``universality classes''. This crucial question
is connected to the predictive power of simplified models and to the 
possibility of applying theoretical results to real low-dimensional
materials.

\subsection{Methods}

Various theoretical approaches to account for the observed phenomenology 
have been developed and implemented. In the rest of the volume they 
will be exposed in detail; here we limit ourselves to a brief description.
The methods discussed are
\begin{enumerate}

\item \textit{Fluctuating hydrodynamics} approach: here the models are described
in terms of the random fields of deviations of the conserved quantities 
with respect to their stationary values. The role of fluctuations
is taken into account by renormalization group or some kind of 
self-consistent theory. 

\item \textit{Mode-coupling} theory: this is closely related to
the above, as it amounts to solving (self-consistently) some approximate equations
for the correlation functions of the fluctuating random fields.  

\item \textit{Kinetic theory}: it is based on the familiar 
approach to phonon transport by means of the Boltzmann equation.
 
\item \textit{Exact solution} of specific models: typically in this case
the original microscopic Hamiltonian dynamics is replaced by 
some suitable stochastic one which can be treated by probabilistic
methods.  

\end{enumerate} 

A sound theoretical basis for the idea that the above described anomalies 
are generic and universal for all momentum-conserving system was put forward in ~\cite{Narayan02}. 
The authors treated the case of a fluctuating $d$-dimensional fluid
and applied renormalization group techniques to evaluate the contribution
of noisy terms to transport coefficients. The calculation 
predicts that the thermal conductivity exponent is $\alpha = (2-d)/(2+d)$. 
From the arguments exposed above, it follows that in 1D the exponents are
\begin{equation}
\alpha \;=\; \delta \;=\; \frac13, 
\qquad \beta =\frac43,  \qquad z \;=\; \frac32 \quad.
\label{13}
\end{equation}
According to this approach, any possible additional term in the noisy
Navier-Stokes equation yields irrelevant corrections in the renormalization
procedure, meaning that the above exponents are model independent, provided
the basic conservation laws are respected. 

Next we give a flavour of one of the other approaches:  the 
Mode-Coupling Theory (MCT). This type of theories has been traditionally invoked
to estimate long-time tails of fluids \cite{Pomeau1975} and to describe the glass
transition \cite{Schilling2005}. In the simplest version, it involves
the normalized correlator of the particle displacement (see Eq.(\ref{dens}),
where the discrete wavenumber $k$ has been turned to the continous
variable $q$)
\[
G(q,t)= \frac{\langle U^*(q,t)U(q,0) \rangle}{\langle
|U(q)|^2 \rangle}  \; .
\] 
$G(q,t)$ is akin to the density--density 
correlator, an observable routinely used in condensed--matter physics.  
The main idea is to write  
a set of approximate equations for $G(q,t)$ that must be solved 
self--consistently.
For the problem at hand, the simplest version of the theory amounts to
consider the equations ~\cite{Scheipers1997,Lepri98c}
\begin{equation}
{\ddot G} (q,t) + 
\varepsilon \int_0^t \Gamma (q,t-s) {\dot G}(q,s) \, ds 
+ {\omega}^2(q) G(q,t)  
= 0 \quad,
\label{mct}
\end{equation}
where the memory kernel $\Gamma(q,t)$ is proportional to $\langle
\mathcal{F}(q,t)\mathcal{F}(q,0) \rangle$, with $\mathcal{F}(q)$ being the
nonlinear part of the fluctuating force between particles. 
Eq.~(\ref{mct}) is derived within the well--known
Mori--Zwanzig projection approach \cite{Kubo1991}.  It must be
solved with the initial conditions $G(q,0)=1$  and $\dot G(q,0)=0$. 

The mode--coupling approach basically amounts to replacing the exact memory
function $\Gamma$ with an approximate one, where higher--orders correlators are
written in terms of $G(q,t)$. 
In the generic case, in which $k_3$ is different from zero (see Eq.~(\ref{fpu})),
the lowest-order mode coupling approximation of the memory kernel turns
out to be ~\cite{Scheipers1997,Lepri98c} 
\begin{equation}
\Gamma(q,t)= \,\omega^{2}(q)
\,\frac{2 \pi}{N} \sum_{p+p'-q=0,\pm\pi}  \,G(p,t) G(p',t) \quad .
\label{mct2}
\end{equation}
Here $p$ and $p'$ range over the whole Brillouin zone (from $-\pi$
to $\pi$ in our units)~. This yields a closed system of nonlinear
integro--differential equations. Both the coupling constant
$\varepsilon$ and the frequency $\omega(q)$ are temperature-dependent input
parameters, which should be computed independently by numerical simulations or
approximate analytical estimates. For the present purposes it is sufficient to 
restrict ourselves to considering their bare values, obtained in
the harmonic approximation. In the adopted dimensionless units they read
$\varepsilon = {3k_3^2 k_BT / 2\pi}$ and  $\omega(q)=2 | \sin\frac{q}{2}|$. Of
course, the actual renormalized values are needed for a quantitative comparison
with specific models. The long-time behaviour of $G$ can be determined by 
looking for a solution of 
the form 
\begin{equation}
G(q,t) \;=\; C(q,t )e^{i \omega (q)t} + c.c. \quad .
\label{g}
\end{equation}
with $\dot G \ll \omega G$. It can thus be shown \cite{Delfini06,Delfini07b} that, 
for small $q$-values and long times $C(q,t) = g(\sqrt{\varepsilon}t q^{3/2})$
i.e. $z=3/2$ in agreement with the above mentioned numerics. 
Furthermore, in the limit $\sqrt{\varepsilon}t q^{3/2} \to 0$ one can
explicitly evaluate the functional form of $g$, obtaining
\begin{equation}
C(q,t) \;=\; \frac12 \exp\left( -D q^2 |t|^{\frac43} \right) \quad,
\end{equation}
where $D$ is a suitable constant of order unity.
The correlation displays a ``compressed exponential'' behaviour 
in this time
range. This also means that the lineshapes of the structure factors
$S(q,\omega)$ are non-Lorenzian but rather exhibit an unusual faster power-law
decay $(\omega - \omega_{max})^{-7/3}$ around their maximum.  
Upon inserting this scaling result into the definition of the heat flux, one
eventually concludes that the conductivity exponent is
$\alpha=1/3$, in agreement with (\ref{13}).

A more refined theory requires considering the mutual interaction 
among \textit{all} the hydrodynamic modes associated with the conservation
laws of the system at hand. The resulting calculations are considerably
more complicated but they can be worked out \cite{VanBeijeren2012,Spohn2014}. 
As a result, the same values of the scaling exponents are found, but also
a more comprehensive understanding is achieved (see Chapter 3 for a detailed 
account). 

\subsection{Connection with the interface problem}

Relevant theoretical insight comes from the link
with one of the most important equations in nonequilibrium statistical 
physics, the Kardar-Parisi-Zhang (KPZ) equation. This is a nonlinear
stochastic Langevin equation which was originally introduced
in the (seemingly unrelated) context of surface growth \cite{Barabasi1995}. 
Let us first consider the fluctuating Burgers equation for the random field
$\rho(x,t)$
\begin{equation}
\frac{\partial \rho}{\partial t}=\frac{\lambda} 2\frac{\partial \rho^2}{\partial x} + 
D\frac{\partial^2 \rho}{\partial x^2} + \frac{\partial \eta}{\partial x},
\label{burgers}
\end{equation}
where $\eta(x,t)$ represents a Gaussian white noise with 
$\langle\eta(x,t)\eta(x',t')\rangle$=$2D\delta(x-x')\delta(t-t')$. 
As it is well-know, Eq.~(\ref{burgers}) can be transformed into the KPZ equation 
by introducing the "height function" $h$ such that $\rho=\frac{\partial h}{\partial x}$, 
\begin{equation}
\frac{\partial h}{\partial t}=\frac{\kappa} 2\left(\frac{\partial h}{\partial x}\right)^2 +\frac D{A}\frac{\partial^2 h}{\partial x^2} +  \eta.
\label{kpzh}
\end{equation}
It has been shown \cite{VanBeijeren2012} that the mode-coupling approximation for 
the correlator of $\rho$ obeying (\ref{burgers}) is basically identical to the 
equation for $C$ described in the previous paragraph. Thus one may argue that the
dynamical properties are those of the KPZ equation in one dimension.
Loosely speaking, we can represent the displacement field as the superposition 
of counterpropagating plane waves modulated by an envelope that is ruled, 
at large scales, by Eq.~(\ref{kpzh}).

\begin{figure}[ht]
\begin{center}
\includegraphics[width=0.75\textwidth,clip]{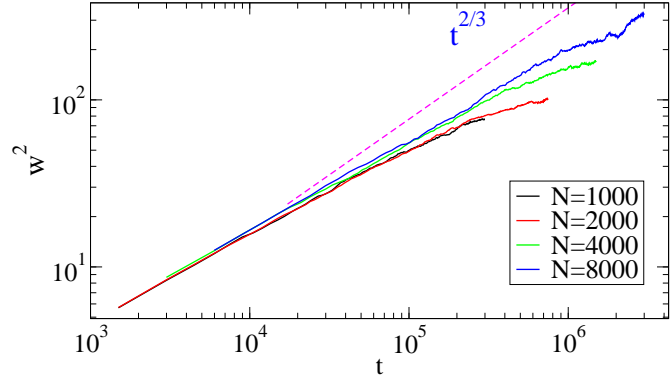}
\begin{center}
\caption{Evolution of the the variance (\ref{hwidth}) for the FPU
$\alpha\beta$ chain with $e=0.5$ $\alpha=0.1$ and different chain lengths.
The dashed line is the expected KPZ growth rate. }
\end{center}
\label{widkpz}
\end{center}
\end{figure}

In order to illustrate this, we have performed a typical 
``KPZ numerical experiment" \cite{Barabasi1995} for the for the FPU
$\alpha\beta$ chain. In practice, we monitored
\begin{equation}
w^2(t,N) = \left\langle \frac1N \sum_n h_n^2 - (\frac1N \sum_n h_n)^2 \right\rangle
\label{hwidth}
\end{equation}
where $h_n(t)=q_n(t)-q_n(0)$, $q_n(0)$ is an equilibrium configuration and the angular
brackets denote an average over an ensemble of different trajectories.
The results are reported in Fig.~\ref{widkpz}.
The only difference with respect to the usual setup is that here the square-width
is plotted only at times $t$ multiples of $L/c$, where $c$ is the effective sound speed.
These are the only moments, when the effect of counterpropagating sound waves cancel out, 
offering the chance to identify a KPZ-like behavior.
In fact, one can see that the growth in Fig.~\ref{widkpz} is compatible with the expected 
KPZ exponent 2/3 (actually, a bit smaller) 
followed by a saturation due to the finite size of the chain.
A more rigorous discussion of the above topics can be found in Chapter 3.

\subsection{Other universality classes}

In the previous section we argued that the scaling properties of anomalous transport
are independent of the microscopic details and correspond to those of the KPZ universality
class. One might wonder whether other classes exist and under which conditions they
can be observed. A reasonable argument, that can be invoked to delimit the KPZ universality 
class, is the \textit{symmetry} of the interaction potential with respect to the 
equilibrium position. With reference to the MCT, one realizes that the symmetry of the fluctuations 
implies that the quadratic kernel in (\ref{mct2}) should be replaced by a cubic 
one\footnote{In fact, the quadratic kernel corresponds to a quadratic force originating from
the leading cubic nonlinearity of any asymmetric interaction potential, 
while a quartic leading nonlinearity of a symmetric interaction potential yields a cubic kernel (force).}, 
thus yielding different values of the exponents \cite{Delfini07b}. In the language of KPZ interfaces,
whenever the coefficient of the nonlinear term vanishes, the evolution equation reduces to the
Edwards-Wilkinson equation that is indeed characterized by different scaling exponents.
The argument can be made more precise in the framework of the full hydrodynamic theory \cite{VanBeijeren2012,Spohn2014}.
There, different dynamical exponents can arise if the coupling between some
modes vanishes (we refer again the reader to Chapter 3 for a detailed discussion).  
A thermodynamic interpretation of this difference is given in \cite{Lee-Dadswell05,Lee-Dadswell08}. 
 
The FPU model is a natural instance to test this working hypothesis. In fact,
systematically larger values of the scaling exponent $\alpha$ have been  reported
for the FPU-$\beta$ case where the cubic term of the potential is absent \cite{Lepri03}.  
The existence of two universality classes for thermal transport in one-dimensional oscillator systems has been also demonstrated in \cite{Lee-Dadswell2015}, where it 
was further proposed that the criterion for being out of the KPZ class 
is the condition $\gamma=c_P/c_V=1$, where $c_P$ and $c_V$ are the specific heat
capacities at constant pressure and volume, respctively.

The scenario can be further illustrated by considering a modification of the HPG model,
the so-called Hard-Point Chain (HPC) \cite{Delfini07a},  characterized by a
square--well potential in the relative distances 
\begin{equation}
 V(y)= \left \{ \begin{array}{ll}
     0 & \quad \mbox{$0\;<y\;<\;a$}\\
     \infty & \quad \mbox{otherwise}
     \end{array} \right.  \; .
\label{u}     
\end{equation}
The infinite barriers at $y=a$ imply an 
elastic ``rebounding" of particles as if they were linked by an inextensible and massless string of fixed
length $a$. The string has no effect on the motion, unless it reaches its maximal length, 
when it exerts a restoring force that tends to rebound the particles one 
against the other. The potential (\ref{u}) introduces the physical 
distance $a$ as a parameter of the model.

%
%
%
%
As it is well known, the thermodynamics of models like the HPC can be solved
exactly and the equation of state is found to be
\[
L =  N \left [ \frac{1}{\beta P}- \frac{a}{\exp(\beta Pa)-1} \right ]
\label{eqst} 
\] 
where $P$ is the pressure of the HPC.  Note that, for
large values of $a$, the equation of state is the same of an HPC 
i.e. the one of an ideal gas in 1D. The important point here is that we can
choose the parameter $a$ such as $P=0$. In this particular point 
the interaction is symmetric ($L/N = a/2$).

A peculiarity of the HPC model is that energy transfer occurs also at 
rebounding ``collisions'' at  distance $a$, this means that besides
the contribution defined by Eq.~(\ref{lj612}) one should include
a term $j'_i$ as from Eq.~(\ref{jsolid}). However, one cannot proceed
directly, since the force is singular, there.
By defining the force between two particles as the momentum difference induced by a collision,
$j'_i$ can be written as the kinetic energy variation times the actual
distance $a$, i.e.  $j'_i = a m_i(u_{i}^{'2}-u_{i}^2)/2$,  divided by a
suitable time-interval $\Delta t$. In order to get rid of the microscopic
fluctuations, it is necessary to consider a sufficiently long $\Delta t$, so
as to include a large number of collisions. Since the number of collisions is
proportional to the system size, it is only in long systems that fluctuations
can be removed without spoiling the slow dynamics of the heat flux. 
Equilibrium simulations show that for $L/N=a/2$ the leading contribution to
the heat flux is given by the term $j'_i$ which exhibits a low-frequency divergence 
with an exponent $\delta = 0.45$, that is not only definitely larger than 1/3
(the value predicted for the KPZ class), but is also fairly close to the
results found for the FPU-$\beta$ model \cite{Lepri03}.  

In out--of--equilibrium simulations, a compatible exponent $\alpha=0.4$ has been measured \cite{Politi2011}.
Those values should be compared with $\alpha=1/2$, the  prediction of mode-coupling theory, 
thus  supporting  the conjecture that the case $P=0$  belongs to a universality
class different from KPZ.


To conclude this Section, let us mention that further support to this
scenario comes from a stochastic model of a chain of harmonic oscillators,  
subject to momentum and energy-conserving noise \cite{BBO06}.
Indeed, one can prove that the dynamical exponents are different
from the KPZ class, e.g. $\delta=1/2$ \cite{BBO06} and $\alpha=1/2$ \cite{Lepri2009}.
Details about this class of models can be found in Chapter 5.
The qualitative explanation is that, as the stochastic collisions 
occur independently of the actual positions,
the effective interaction among particles is symmetric and thus
equivalent to the $P=0$ case. Notably, this remains true 
even if the harmonic potential is replaced by an anharmonic one, like 
the FPU-$\alpha\beta$ \cite{Basile08}.
Finally, the application of kinetic theories to the
$\beta$-FPU model \cite{Pereverzev2003,Nickel07,Lukkarinen2008} yields
a non-KPZ behavior, $\alpha=2/5$. We refer the reader to Chapter 4 for 
a detailed account.

\subsection{Comparison with simulations and experiments}

The theoretical predictions have been intensively investigated in the 
recent literature. A direct validation by numerical simulations is, to some 
extent, challenging and has been debated through the years \cite{Delfini08c}. 
Generally speaking, the available numerical estimates of $\alpha$ and $\delta$ 
may range between 0.25 and 0.44~\cite{LLP03}. As a matter 
of fact, even in the most favorable cases of
computationally efficient models as the HPG, finite--size corrections to 
scaling are sizeable. In this case, $\alpha$--values as diverse as 0.33
~\cite{Grassberger02} and 0.25~\cite{Casati2003} for comparable parameter choices have been
reported. On the other hand, a numerically convincing confirmation of the 
$\alpha=1/3$ prediction comes from the diffusion of perturbations \cite{Cipriani05}.
We refer to Chapter 6 for some detailed numerics.

The ultimate goal would be of course the validation of the universality hypothesis 
in more realistic systems, possibly characterized by more than one degree of freedom
per lattice site.  The first
remarkable attempt was the application to the vibrational dynamics 
of individual single--walled carbon nanotubes, which can be 
in many respect considered as one-dimensional objects. Signature
of anomalous thermal transport was first reported in molecular dynamics
simulations in \cite{Maruyama2002}. Note that this type of simulations involve 
complicated three-body interactions among carbon atoms, 
thus supporting the claim that toy models like ours can indeed capture 
some general features.  We refer the reader to Chapter 7 for a critical
discussion of molecular dynamics results on carbon-based material. 
Chapter 8 will report some experimental data on nanotubes and 
nanowires and discuss the current state of the art.

%
%
%
%

\section{The coupled rotors model}
\label{sec:rotors}

As discussed in the previous Sections, one-dimensional anharmonic chains generically display 
anomalous transport properties. A prominent exception is the 
coupled rotors chain described by the equation of motion
\begin{equation}\label{2}
\dot q_n = p_n\,,\quad \dot p_n = \sin(q_{n+1} - q_n)- \sin(q_{n} - q_{n-1})\,.
\end{equation}
The model is sometime referred to as the Hamiltonian version of the XY spin chain. 
The energy flux is $j^e_n= \langle p_n\sin(q_{n+1} - q_n)\rangle$.
As the interaction depends only on the angle differences, 
angular momentum is conserved and one may expect anomalous transport to
occur. Nevertheless, molecular dynamics 
simulations have convincingly demonstrated 
normal diffusion \cite{Giardina99,Gendelman2000,Yang2005}. 

There are two complementary views to account for this difference. In the general
perspective of nonlinear fluctuating hydrodynamics, the chain ``length" $L$
defined as $L=\sum_n (q_{n+1}-q_n)$ is not even well defined,
because of the phase slips of $\pm 2\pi$, so the corresponding evolution 
equation breaks down and normal transport is eventually expected.
From a dynamical point of view, one can invoke that normal transport sets in due to the
spontaneous formation of local excitations, the so--called  
\textit{rotobreathers}, that behave like scattering centers \cite{Flach2003}. 
Phase slips (jumps over the energy barrier), on their side,  may  effectively
act as localized random kicks, that contribute to scatter the 
low-frequency modes, thus leading to a finite conductivity. In order to test
the validity of this conjecture, one can study the temperature dependence of 
$\kappa$ for low temperatures $T$, when jumps across barriers become increasingly
rare. Numerics indicates that the thermal conductivity behaves as 
$\kappa \approx \exp (\eta/T)$ with $\eta \approx 1.2$. The same  kind of dependence on $T$
(although with $\eta \approx 2$) is found for the average
escape time $\tau$ across the potential barrier: this can be explained by 
assuming that the phase slips are the results of activation processes. 

An important extension is the 2D case, i.e. rotors coupled to their 
neighbors on a square lattice, akin to the celebrated XY--model. As it is well known, the latter is
characterized by the presence of the so called Kosterlitz-Thouless-Berezinskii
phase transition at a temperature $T_{KTB}$, between a disordered
high--temperature phase and a low--temperature one, where vortices condensate.
It is likely that transport properties are qualitatively different
in the two phases. Numerical simulations \cite{Delfini2005} performed on a 
finite lattice indeed 
show that  they are drastically different in the high--temperature and in the
low--temperature phases. In particular, thermal conductivity is finite in the
former case, while in the latter it does not converge up to lattice sizes of
order 10$^4$. In the region where vorticity is negligible ($T < 0.5$) the
available data suggest a logarithmic divergence with the system size, analogous
to the one observed for coupled oscillators (see next Section).  Close to $T_{KBT}$,
where a sizable density of bounded vortex pairs are thermally excited, numerical
data still suggests a divergence, but the precise law has not be reliably estimated.

\section{Two-dimensional lattices}
\label{sec:2D}

Heat conduction in $2D$ models of anharmonic oscillators coupled through 
momentum--conserving interactions is expected to exhibit different properties
from those of $1D$ systems. In fact, extension of the arguments discussed in
the previous sections predicts a logarithmic divergence of
$\kappa$ with the system size $N$ at variance with the power--law predicted
for the $1D$ case. Consideration of this case is not only for completeness
of the theoretical framework, but is also of great interest for 
almost-2D  materials, like graphene, that will be treated in the 
Chapters 7 and 9.

Although the theory in this case if far less developed,
there are several numerical evidences in favor of such logarithmic divergence.
In \cite{Lippi00}, a square lattice of oscillators interacting through the
FPU-$\beta$  (see Eq.~(\ref{fpu}), with $k_3=0$) or the Lennard-Jones 
(see Eq.~(\ref{lj})) potentials, was investigated by means of both   
equilibrium and nonequilibrim simulations.  The models are
formulated in terms of two-dimensional vector displacements $u_{ij}$ 
and velocities and $\dot u_{ij}$, defined on a square 
lattice containing $N_x \times N_y$ atoms of equal masses $m$ and  
nearest-neighbor interactions. Periodic and fixed boundary conditions have been 
adopted in the direction perpendicular ($y$) and parallel ($x$) to the thermal gradient, respectively.
Simulations for different lattice sizes have been performed by keeping the 
ratio $N_y/N_x$ constant and not too small to observe genuine $2D$ features
(in \cite{Lippi00} $ N_y/N_x =1/2$ was chosen).

The simulations reveal several hallmarks of anomalous behavior:
temperature profiles display deviations from the linear shape predicted by Fourier law
and the size dependence of the thermal conductivity is well-fitted by a logarithmic law
\begin{equation}
\kappa = A + B \log N_x \; ,
\label{cond2d}
\end{equation}
with $A$ and $B$ being two unknown constants. A consistent indication
comes from the evaluation of the Green-Kubo integrand in the microcanonical
ensemble. Indeed, the energy-current autocorrelation is compatible
with a decay $1/t$ at large times.

Despite these first indications, the numerics turns out to be very difficult,
which is not surprising in view of the very weak form of the anomaly, peculiar
of the 2D case.  As a matter of fact very robust finite-size effects are observed in the calculations 
for other lattices, which well exemplify the difficulties in observing the true asymptotic 
behavior with affordable computational resources \cite{Wang2012}.

Another interesting issue concerns dimensional crossover, namely how  
the divergence law of the thermal conductivity will change from the 2D class to 1D 
class as $N_y/N_x$ decreases. This issue has been studied 
for the two-dimensional FPU lattice in \cite{Yang06}. 
We refer to Chapter 6 for a further detailed discussion.

\section{Integrable nonlinear systems}
\label{sec:integrable}

The harmonic crystal behaves as an ideal conductor, because its dynamics can be
decomposed into the superposition of independent ``channels''. This peculiarity
can be generalized to the broader context of integrable nonlinear systems.
They are mostly one-dimensional models characterized by the presence of 
``mathematical solitons", whose stability is determined by the interplay of 
dispersion and nonlinearity. This interplay is expressed by the existence of a 
macroscopic number of {\it conservation laws}, constraining the dynamical 
evolution. Intuitively, the existence of freely
travelling solitons is expected to yield ballistic transport, i.e. an
infinite conductivity. From the point of view of the Green-Kubo formula, this
ideal conducting behavior is reflected by the existence of a nonzero flux
autocorrelation at arbitrarily large times. This, in turn, implies that the
finite-size conductivity \textit{diverges linearly with the system size}.

Although integrable models are, in principle, exactly solvable, the actual
computation of dynamic correlations is technically involved. A more
straightforward approach is nevertheless available to evaluate the 
asymptotic value of the current autocorrelation. This is accomplished by means 
of an inequality due to Mazur \cite{Mazur1969} that, for a generic observable 
$\mathcal A$, 
(with $\langle \mathcal A\rangle=0$, where $\langle \ldots \rangle$ denotes 
the equilibrium thermodynamic average) reads 
\begin{equation}
\lim_{\tau \rightarrow \infty} \frac{1}{\tau} \int_0^ \tau 
\langle \mathcal A(t)\mathcal A(0)\rangle  \, dt \; \geq \; \sum_n
\frac{\langle \mathcal A \mathcal Q_n\rangle^2}{\langle \mathcal Q_n^2\rangle} \quad,
\label{mazur}
\end{equation}
where ${\mathcal Q_n}$ denote a set of conserved and mutually orthogonal 
quantities, ($\langle \mathcal Q_n \mathcal 
Q_m\rangle=\langle \mathcal Q_n^2\rangle\delta_{n,m}$).

In the present context the most relevant example is the equal-masses Toda chain 
with periodic boundary conditions, defined, in reduced units, by the
Hamiltonian
\begin{equation}
H \;=\; \sum_{n=1}^N \left[ \frac{p_n^2}{2}+\exp(-r_n) \right] \; ,
\label{toda}
\end{equation}
where $r_n=q_{n+1}-q_n$ is the relative position of neighboring particles. 
The model is completely integrable, since it admits $N$ independent constants
of the motion \cite{Henon1974,Flaschka1974}.
Lower bounds on the long time value of $\langle J(t)J(0)\rangle$ can be
calculated through the inequality (\ref{mazur}) \cite{Zotos02}. The resulting lower bound
to the conductivity is found to increase monotonously with the temperature. At
low $T$, the growth is linear with a slope comparable to the density of solitons
$N_s/N=(\ln 2/\pi^2) T$. This trend is interpreted as an evidence for the
increasing contribution of thermally excited nonlinear modes to ballistic
transport.

To conclude, let us also mention that Mazur-type of inequalities have 
been recently used as a theoretical basis for the study of thermoelectric 
coefficients. This is discussed in Chapter 10 of the present volume.

\section{Coupled transport}
\label{sec:coupled-transport}

Up to this point we have restricted the discussion to models where just one
quantity, the energy, is exchanged with external reservoirs and transported across the system.
In general, however, the dynamics can be characterized by more than one conserved quantity.
In such cases, it is natural to expect the emergence of coupled transport phenomena, in the sense 
of ordinary linear irreversible thermodynamics. Works on this problem
are relatively scarce~\cite{Gillan85,Mejia2001,Larralde03,Basko2011}. Interest in
this field has been revived by recent works on thermoelectric
phenomena~\cite{Casati2008,Casati2009,Saito2010} in the hope of identifying dynamical
mechanisms that could enhance the efficiency of thermoelectric energy
conversion. This will be treated in detail in Chapter 10.

Here, we briefly discuss two models: a chain of coupled rotors and the discrete nonlinear
Schr\"odinger equation, where the second conserved quantity is the momentum and
the norm (number of particles), respectively.

\subsection{Coupled rotors}

The evolution equation defined in (\ref{2}) must be augmented to include
the exchange of momentum with the external reservoirs,
\begin{eqnarray}
  \dot{p}_n &=& \sin(q_{n+1}-q_n) - \sin(q_n-q_{n-1})+\\
 && \delta_{1n}\left (\gamma (F_+-p_1) + \sqrt{2\gamma T_+}\,\eta_+\right) +
 \delta_{1N}\left (\gamma (F_--p_N) + \sqrt{2\gamma T_-}\,\eta_-\right)  \nonumber
 \label{xymodel}
\end{eqnarray}
where $F_\pm$ and $T_\pm$ denote the torque applied to the chain boundary and
the corresponding temperature, respectively; $\gamma$ is the coupling strength with the 
external baths and $\eta_\pm$ is a Gaussian white noise with unit variance.
The effect of external forces on the Hamiltonian XY model has been preliminarly addressed in
~\cite{Eleftheriou2005,Iacobucci2011, Iubini2014}. 

As discussed in section \ref{sec:rotors}, (angular) momentum is conserved and one can, in fact,
define the corresponding flux as $j^p_n= \sin(q_{n+1}-q_n)$.
A chain of rotors is perhaps the simplest model where one can exert a gradient of forces 
that couples to heat transport, giving rise to nontrivial phenomena, even
though the transport itself is normal.
For  $F_+=F_-$, all the oscillators rotate with the same frequency $\omega= F$, no
matter which force is applied: no momentum flux is generated. 
In fact, what matters is the difference between the forces applied at the two extrema
of the chain. Therefore, from now on we consider the case of  zero-average force, i.e. 
$F_+=-F_-$. In the presence of such a gradient of forces, the oscillators may rotate
with different frequencies and, as a result, a coupling between angular momentum
and energy transport may set in.
In principle, one could discuss the same setup for general chains of kinetic
oscillators, as (linear) momentum is conserved in that context too.
However, nothing interesting is expected to arise. For a binding
potential, like in the FPU model, the presence of an external force is akin
to the introduction of a homogeneous, either positive or negative, pressure
all along the chain. In fact, the pressure $P$ is, by definition, equal to the equilibrium average 
of the momentum flux, $P= \langle j^p \rangle$ ( at equilibrium, the r.h.s. is independent
of $n$) .  
On the other hand, if the potential is not binding (e.g., the  Lennard-Jones chain (\ref{lj})) and the applied
force is equivalent to a negative pressure, the system would break apart.

In the presence of two fluxes, the linear response theory implies that they must satisfy
the equations \cite{Saito2010} 
(angular brackets denote an ensemble, or equivalently, a time average, assuming ergodicity)
\begin{eqnarray}
\label{lin}
\langle j^p \rangle &=& -L_{pp} \frac{d (\beta\mu)}{dy} + L_{pe} \frac{ d \beta}{dy} \\
\langle j^e\rangle &=& -L_{ep} \frac{d (\beta\mu)}{dy} + L_{ee} \frac{ d \beta}{dy} \; ,\nonumber
\end{eqnarray}
where $y=n/N$, $\beta$ is the inverse temperature $1/T$ (in units of the Boltzmann constant)
and $\mu$ is the chemical
potential, which, in the case of the coupled rotors, coincides with the
average angular frequency $\omega_n =\langle p_n \rangle$. Finally,
$\bf{L}$ is the symmetric, positive definite, $2\times 2$ Onsager matrix. If $L_{ep}=0$, the
two transport processes are uncoupled. 

\begin{figure}[ht]
\begin{center}
\includegraphics[width=0.75\textwidth,clip]{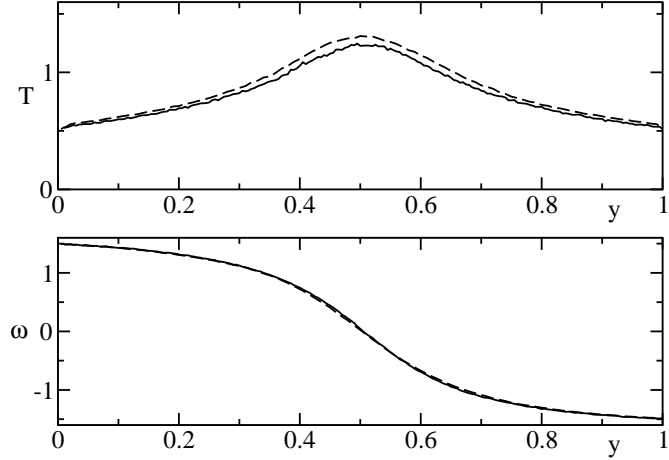}
\caption{Stationary profile of the temperature (upper panel) and of the average frequency (lower panel)
for $T(0)=T(1)=0.5$, $F=1.5$, and $\gamma=1$; $y=n/N$. The dashed and solid curves correspond to
$N=100$, and 200, respectively.}
\label{F.coupled01}
\end{center}
\end{figure}

In the case of the rotor chain, it is important to realize that a
correct definition of the kinetic temperature requires subtracting the coherent
contribution due to the nonzero angular velocity, i.e.
\[
T_n=\langle (p_n-\omega_n)^2\rangle \; .
\]
The effect of coupling between energy and momentum transport is better understood
by considering a setup where the two thermal baths operate at the same temperature $T$.
Because of the flux of momentum, the temperature profile deviates
from the value imposed at the boundaries.
In Fig.~\ref{F.coupled01} we show the results for $T= 0.5$ and $F=1.5$ and two different system sizes.
Notably, the temperature profile displays a peak in the central region 
\cite{Iacobucci2011}, where it reaches a value
around 1.2; the average frequency varies nonuniformly across the sample with a steep
region in correspondence of the central hot spot. At the same time, the energy flux
$j^e$ is zero, so that the anomalous behaviour of the temperature profile is
entirely due to the coupling with the nonzero momentum flux.

\begin{figure}[htbp]
\begin{center}
\includegraphics[width=0.75\textwidth,clip]{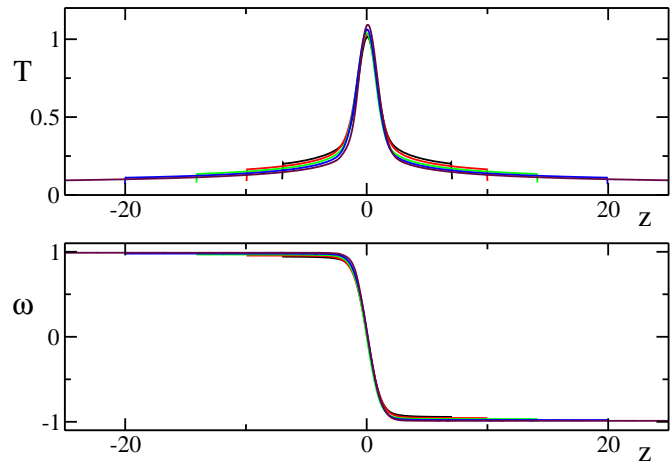}
\caption{Stationary profile of the temperature (upper panel) and of the average frequency (lower panel)
for $T(0)=T(1)=0.$, $F=1.05$, and $\gamma=1$; $z=(n-N/2)/N^{1/2}$. The various curves correspond
to $N=200$, 400, 800, 1600, and 3200.}
\label{F.profiles}
\end{center}
\end{figure}

This behaviour can be traced back to the existence of a (zero-temperature) boundary-induced transition.
In fact, for $T=0$, there exists a critical torsion $F_c=1/\gamma$ \cite{Iubini2014} such that
for $F<F_c$ the ground state is a twisted fully-synchronized state, whereby each element is
at rest and is characterized by a constant phase gradient. Here, $T_n=0$ throughout the whole lattice.
For $F>F_c$ the fully sinchronized state turns into a chaotic asynchronous dynamics
with $\omega_1 = F = -\omega_N$. Remarkably, even though both heat baths operate at
zero temperature and the equations are deterministic and dissipative, the temperature in the middle
raises to a finite value (see Fig.~\ref{F.profiles})  even in the thermodynamic limit.

The phenomenon can be interpreted as the onset of an interface (the hot region) separating two different
phases: the oscillators rotating with a frequency $F$ (on the left) from those rotating
with a frequency $-F$ (on the right). The phenomenon is all the way more interesting in view
of the anomalous scaling of the interface width with the system size (it grows as $N^{1/2}$,
see Fig.~\ref{F.profiles}) and its robustenss (it is indepedent of the value of the torsion  $F$, provided it
is larger than the critical value $F_c$ \cite{Iubini2014}).

Accordingly, the interface is neither characterized by a finite width nor it is extensive.
A more careful inspection reveals that the $N^{1/2}$ width is due to a spatial
Brownian-like behavior of an instantaneously much thinner interface.
Nevertheless, even the instantaneous interface extends over a diverging number of sites, 
of order $N^{1/5}$, thus leaving the anomaly fully in place.
Such a state can neither be predicted within a linear-response type of
theory, nor traced back to some underlying equilibrium transition. Even more remarkably, it constitutes 
an example of a highly inhomogeneous, unusual chaotic regime. 
Indeed, while the fractal dimension is extensive (i.e. proportional to the number of oscillators)
the Kolmogorov-Sinai (KS) entropy is not: it increases only as $N^{1/2}$.
The KS entropy measures the diversity of the ``ground state" non-equilibrium configurations
that are compatible with the given thermal baths. Its lower-than-linear increase with $N$
implies that we are not in the presence of a macroscopic degeneracy, as in spin glasses. 

The anomaly of the regime is finally reinforced by the scaling behavior of the momentum
flux, which scales as $N^{-1/5}$. A theoretical explanation of this behavior is still
missing. All of these anomalies disappear as soon as the temperature at the boundaries
is selected to be strictly larger than zero.
In particular, the width of the hot spot suddenly becomes extensive and the scaling
of the momentum is normal ($j^p \simeq 1/N$). The nonmonotonous behavior of the
temperature is nevertheless a nontrivial consequence of the coupling between
heat and momentum transport.

\subsection{The Discrete Nonlinear Schr\"odinger equation} 

The above discussed non--equilibrium transition is not
a peculiarity of the rotor model. A similar scenario can be
observed also in the Discrete Nonlinear Schr\"odinger (DNLS) equation 
\cite{Eilbeck1985,Kevrekidis}, a model with important applications in many domains of
physics. In one dimension, the DNLS Hamiltonian is 
\begin{equation}
H=\frac{1}{4}\sum_{n=1}^{N}\left(p_n^2+q_n^2\right)^2+
\sum_{n=1}^{N-1}\left(p_np_{n+1}+q_nq_{n+1}\right)
\quad ,
\label{ham}
\end{equation}
where the sum runs over the $N$ sites of the chain. The sign of the quartic term is
positive, while the sign of the hopping term is irrelevant, due to the symmetry 
associated with the canonical (gauge) transformation $z_n \to z_ne^{i\pi n}$ (where
$z_n\equiv (p_n + \imath q_n)/\sqrt{2}$ denotes the amplitude of the wave
function). The equations of motion are
\begin{equation}
i\dot z_n = -z_{n+1}-z_{n-1}-2|z_n|^2 z_n
\label{dnlse} 
\end{equation}
with $n = 1, \cdots, N$, and fixed boundary conditions ($z_0=z_{N+1}=0$).
The model has two conserved quantities, the energy and the total norm (or total
number of particles) 
\begin{equation}
A=\sum_{n=1}^{N}(p_n^2+q_n^2) = \sum_{n=1}^{N} |z_n|^2  \quad ,
\label{norm}
\end{equation} 
so that it is a natural candidate for the study of coupled transport.

Since the Hamiltonian is not the sum of a kinetic and potential energy, the thermal
baths cannot be described by standard Langevin equations. 
An effective strategy has been proposed in Ref.~\cite{Iubini2013a}.  Here below we
report the evolution equation for the first oscillator, in contact with a thermal
bath at temperature $T_+$ and with a chemical potential $\mu_+$ (a similar equation
holds for the last particle at site $N$)
\begin{eqnarray}
\label{olla2}
\hspace{-1.cm} 
\dot p_1&=&-(p_1^2+q_1^2)q_1-q_{2}-\gamma\left[(p_1^2+q_1^2)p_1+p_2 -\mu_+ p_1 \right] 
 +\sqrt{2\gamma T_+} \xi'_1 \\ 
\hspace{-1.cm} \dot q_1&=&(p_1^2+q_1^2)p_1+p_{2} -\gamma\left[(p_1^2+q_1^2)q_1+q_2-\mu_+ q_1\right] 
 +\sqrt{2\gamma T_+} \xi''_1 \nonumber  \quad ,
\end{eqnarray}
where $\gamma$ measures the coupling strength with the thermal bath,
while $\xi'_1$ and $\xi''_1$ define two independent white noises with unit variance.
It can be easily seen that the deterministic components of the thermostat,
are gradient terms. As a result, in the absence of thermal noise, they would drive 
the system towards a state characterized by a minimal $(H-\mu A)$.
Notice the nonlinear structure of the dissipation terms in Eq.~(\ref{olla2}).

An additional problem of the DNLS model is the determination of the temperature,
as one cannot rely on the usual kinetic definition (this is again a consequence
of the nonseparable Hamiltonian). An operative definition can be, however,
given by adopting the microcanonical approach \cite{Rugh1997}, i.e. by invoking the thermodynamic
relationships,
\[
T^{-1}=\frac{\partial{\mathcal S}}{\partial{H}} \quad, \quad  
\frac{\mu}{T}=-\frac{\partial{\mathcal S}}{\partial{A}}\; ,
\]
where $\mathcal S$ is the entropy. As shown in \cite{Franzosi2011b,Iubini2012},
the partial derivative $\partial{\mathcal S}/\partial{C_i}$ ($i=1,2$, with $C_1=H$ and
$C_2=A$) can be computed by exploiting the fact that $C_i$ is a conserved quantity,
\begin{equation}
\frac{\partial \mathcal S}{\partial C_i}=
\left\langle \frac{W\|\vec \xi\|}{\vec\nabla C_i\cdot\vec\xi}\
\vec\nabla\cdot\left(\frac{\vec\xi}{\|\vec\xi\|W} \right)\right\rangle
\label{tDnlse}
\end{equation}
where $\langle \: \rangle$ stands for the microcanonical average,
\begin{eqnarray}
\label{add}
\vec \xi&=&\frac{\vec{\nabla} C_1}{\|\vec{\nabla} C_1\|} -\frac{(\vec{\nabla}
C_1\cdot\vec{\nabla} C_2)\vec{\nabla}
C_2}{\|\vec{\nabla} C_1\|\|\vec{\nabla} C_2\|^2}
\\     
W^2&=&
\sum_{\genfrac{}{}{0pt}{}{m,n=1}{m<n}}
^{2N}\left[\frac{\partial C_1}{\partial x_m}\frac{\partial C_2}{\partial x_m}-
\frac{\partial C_1}{\partial x_n}\frac{\partial C_2}{\partial x_m}\right]^2
 \, , \nonumber
\end{eqnarray}
and $x_{2n}=q_n$, $x_{2n+1}=p_n$. The resulting definitions of $T$ and $\mu$ have the unpleasant
property of being nonlocal: numerical simulations, however, show that they give meaningful
results even when they are implemented for relatively short subchains.

As for the fluxes, they are naturally defined from the continuity equations for energy and norm
\begin{equation}
j_n^e= \dot q_n q_{n-1}+\dot p_n q_{n-1} \qquad j_n^p= q_n p_{n-1}-p_n q_{n-1} \; ,`
\end{equation}
Notice that for the sake of simplicity we still use the same notations as in the previous
setup altough here $j^p_n$ denotes the flux of norm/mass rather than momentum.

\begin{figure}[ht]
\begin{center}
\includegraphics[width=0.7\textwidth,clip]{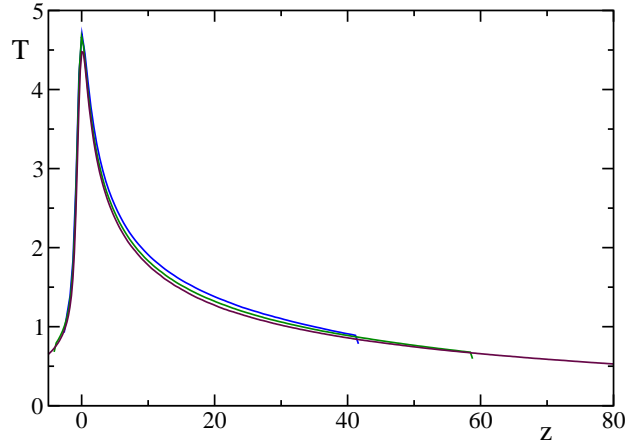}
\caption{Temperature profiles of the DNLS equation for 2000, 4000, 8000 and  
$T=0$, $\mu_+=2$ and $\mu_-=5$; $z = (n-\hat n)/\sqrt{N}$, where $\hat n$ is the site
with the highest temperature.}
\label{F.DNLS}
\end{center}
\end{figure}

If one sets $T_+=T_-=0$, as in the XY model, the control parameter, 
i.e. the driving force, is given by $\delta\mu = |\mu_--\mu_+|/2$ 
\cite{Iubini2013a}. When $\delta\mu$ is larger than a critical value (that here
depends on $A$), a bumpy temperature profile spontaneously emerges. 
As shown in Fig.~\ref{F.DNLS}, the left-right symmetry of the profile found in the XY
model is lost, but the width of the peak still scales as $N^{1/2}$. 
A second crucial difference is the scaling behaviour of the norm-flux, which decreases 
as $N^{-2/5}$ instead of $N^{-1/5}$. This suggests that more than one universality
class is presumably present: the symmetry of the profile might play a crucial role.

In coupled transport, each conservation law implies the presence of a
corresponding thermodynamic variable. In the case of the DNLS equation, there are two
of them: the temperature $T$ (or, equivalently $\beta$) and the chemical potential $\mu$.
If the extrema of a given system are ``attached" to two different points in the $(\mu,T)$ space, 
a new question arises with respect to the transport of just one variable: 
the selection of the path in the phase plane. This problem can be solved with the help
of the linear transport equations (\ref{lin}), which can be rewritten as
\begin{equation}
\label{lin2}
\frac{ d \beta}{d\mu} = \frac{ 
\langle j^e \rangle \beta L_{pp} -
\langle j^p \rangle \beta L_{ep}}
{\langle j^e \rangle \left(L_{pe}-\mu L_{pp}\right) 
- \langle j^p \rangle \left(L_{ee}-\mu L_{ep}\right)} \; .
\end{equation}
The above first order differential equation can be solved once
the Onsager matrix is known across the thermodynamics phase-diagram
and the ratio of the two fluxes is given. This determines unambigiously
the resulting temperature and chemical-potential profiles.

It is worth recalling that in the absence of a mutual
coupling between the two transport processes (zero off-diagonal elements of the
Onsager matrix) such curves would be vertical and horizontal lines in the
latter representation. It is remarkable that the solid
lines, which correspond to $j^e=0$, are almost vertical for large $\mu$:
this means that in spite of a large temperature difference, the energy flux
is very small. This is an indirect but strong evidence that the nondiagonal
terms are far from negligible.

\begin{figure}[htbp]
\begin{center}
\includegraphics[width=0.7\textwidth,clip]{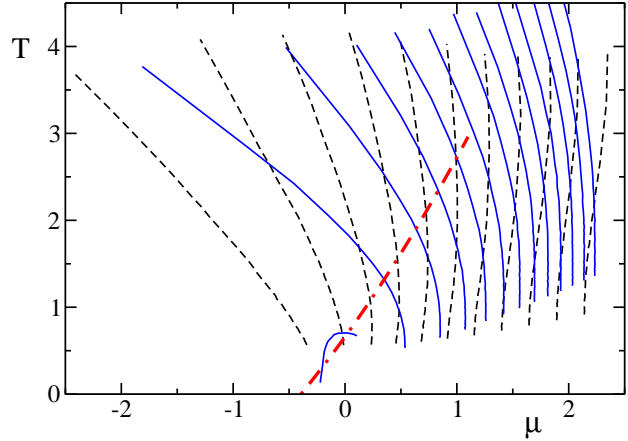}
\caption{Zero-flux curves in the 
$(\mu,T)$ planes.
Black dashed lines correspond to $j^p=0$ and are obtained
with norm-conserving thermostats upon fixing the total norm
density $a_{tot}$, $T_L$ and $T_R$.
Blue solid lines are for $j^e=0$ using energy-conserving
thermostats with fixed total energy density $h_{tot}$, $\mu_L/T_L$
and $\mu_R/T_R$.
Simulations are for a chain of length $N=500$.
The thick dot-dashed lines identify the locus where
$S$ changes sign (see text).}
\label{noflux}
\end{center}
\end{figure}

The condition of a vanishing particle flux $j^p=0$ defines the Seebeck
coefficient which is $S= - d\mu/dT$. Accordingly, the points in Fig.~\ref{noflux},
where the dashed curves are vertical identify the locus where $S$ changes
sign. The $j^e=0$ curves have no direct interpretation in terms of
standard transport coefficients.

\section{Conclusions and open problems}
\label{sec:conclusions}

In the previous sections we have seen that various theoretical approaches
predict the existence of two universality classes for the divergence of
heat conductivity in systems characterized by momentum conservation.
Although this scenario is generally confirmed by numerical simulations, 
some exceptions have been found as well.
The most notable counterexample is the normal conduction which emerges in
chains of coupled rotors. As we have already discussed in section \ref{sec:rotors},
it is quite clear that the peculiarity of this model is to be traced back to
the $2\pi$-slips of the angles $q_n$.

Further, less-understood, anomalies have been found in models where $q_n$ is a genuine
displacement variable. One example is a momentum conserving modification of the famous “ding-a-ling” model. 
The system composed of two kinds of alternating point particles ($A$ and $B$): 
the $A$ particles mutually interact via nearest-neighbour harmonic forces; the $B$ particles
are free to move and collide elastically with the $A$ particles. 
Equilibrium and non-equilibrium numerical simulations indicate that the termal conductivity $\kappa$
is finite \cite{Lee-Dadswell2010}.

Normal heat transport in accordance to Fourier law has been claimed also in simulations
of the FPU-$\alpha\beta$model (and of other asymmetric potentials), at low-enough energies/temperatures
\cite{Zhong2012}. More detailed numerical simulations, however,
indicate that the unexpected results for asymmetric potentials do not represent the 
asymptotic behavior \cite{Wang2013,Das2014}, but rather follow from an insufficient chain length.
This if further strengthened in  \cite{Lee-Dadswell2015a} where 
mode-coupling arguments have been used to determine the frequency below which 
finite-size effects are negligible. It turns out that, in some cases, the asymptotic behavior
may only be seen at exceedingly low frequencies (and thereby exceedingly large system-sizes).

More recent studies report a finite thermal conductivity in the thermodynamic limit 
for potentials that allow for bond dissociation 
(like e.g Lennard-Jones, Morse, and Coulomb potentials) \cite{Savin2014,Gendelman2014}. 
This is explained by invoking phonon scattering on the locally strongly-stretched loose 
interatomic bonds at low temperature and by the many-particle scattering at high temperature. 
On the other hand, the hard-point gas, a model where ``dissociation" arises automatically, without the
need to overcome an energy barrier, was found to exhibit a clean divergence of the conductivity.
Anyway, the universality of scaling in this model has been recently challenged by numerical
studies of the hard--point gas with alternate masses and thermal baths at different temperatures
acting at the boundaries. When the mass ratio is varied, the anomalous exponent is found to
depart significantly from the value 1/3 predicted by the nonlinear fluctuating hydrodynamics
\cite{Hurtado2015}.

Irrespective whether the above discrepancies are a manifestation of strong 
finite-size corrections, or of the existence of another universality classes, 
where the standard hydrodynamic theories do not apply, they have to be explained.

\begin{acknowledgement}
We wish to thank L. Delfini and S. Iubini for their effective contribution to the achievement of
several results summarized in this chapter.

\end{acknowledgement}
%

%
%


\bibliographystyle{spmpsci}
\bibliography{heat,levy,books,diodo}
\end{document}